\documentclass[iop]{emulateapj}
\usepackage{natbib}
\usepackage{amsmath}
\usepackage{graphicx}
\usepackage{txfonts}
\usepackage[scaled]{helvet}
\usepackage{epsfig}
\usepackage[normalem]{ulem}
\usepackage{color}

\begin{document}

\title{Star-disk interactions in multi-band photometric monitoring of the classical T Tauri star GI Tau}


\author{Zhen Guo $^{1,2}$, Gregory J. Herczeg$^1$, Jessy Jose$^1$,
Jianning Fu$^{3}$, Po-Shih Chiang$^{4}$,  Konstantin Grankin$^{5}$, Ra\'{u}l Michel$^{6}$, Ram Kesh Yadav$^{7}$, Jinzhong Liu $^{8}$ Wen-ping Chen$^{4}$, Gang Li$^{3}$, Huifang Xue$^{3}$, Hubiao Niu$^{3,8}$, Annapurni Subramaniam$^{9}$, Saurabh Sharma$^{10}$, Nikom Prasert$^{7}$, Nahiely Flores-Fajardo$^{1,11}$, Angel Castro$^{6,12}$, Liliana Altamirano$^{6,12}$}

\altaffiltext{1}{Kavli Institute for Astronomy and Astrophysics,
  Peking University, Yi He Yuan Lu 5, Haidian District, Beijing 100871,
  P.R. China; guozhen9057@hotmail.com}
\altaffiltext{2}{Department of Astronomy, School of Physics, Peking
  University, Yi He Yuan Lu 5, Haidian District, Beijing 100871,
  P.R. China}
  \altaffiltext{3}{Department of Astronomy, Beijing Normal University, Beijing 100875, P.R. China}
 \altaffiltext{4}{ Graduate Institute of Astronomy, National Central University, 
  No. 300, Zhongda Rd., Zhongli Dist., Taoyuan City 32001, Taiwan}
 \altaffiltext{5}{Crimean Astrophysical Observatory, pos. Nauchnyi, Crimea, 298409 Russia}
 \altaffiltext{6}{Instituto de Astronom\'ia, Universidad Nacional Aut\'onoma de M\'exico. Apartado Postal 877, C.P. 22800, Ensenada, B.C., M\'exico}
 \altaffiltext{7}{National Astronomical Research Institute of Thailand, Chiang Mai, 50200, Thailand}
 \altaffiltext{8}{Xinjiang Astronomical Observatory, Urumqi, Xinjiang 830011, China}
 \altaffiltext{9}{Indian Institute of Astrophysics, Koramangala, Bangalore  560 034, India}
 \altaffiltext{10}{Aryabhatta Research Institute of Observational Sciences, Manora Peak, Nainital 263 002, India}
 \altaffiltext{11}{Instituto de Ciencias Nucleares, Universidad Nacional Aut\'onoma de M\'exico. Cd. Universitaria, 04510 Ciudad de M\'exico, M\'exico}
 \altaffiltext{12}{Universidad Aut\'onoma de Ciudad Ju\'arez, Instituto de Ingenier\'ia y Tecnolog\'ia. 1210 Plutarco El\'ias Calles, 32310 Cd. Ju\'arez, CH, M\'exico}

\begin{abstract}

The variability of young stellar objects is mostly driven by star-disk interactions. In long-term photometric monitoring of the accreting T Tauri star GI Tau, we detect extinction events with typical depths of $\Delta V \sim 2.5$ mag that last for days-to-months and often appear to occur stochastically.  In 2014 -- 2015, extinctions that repeated with a quasi-period of 21 days over several months is the first empirical evidence of slow warps predicted from MHD simulations to form at a few stellar radii away from the central star.  The reddening is consistent with $R_V=3.85\pm0.5$ and, along with an absence of diffuse interstellar bands, indicates that some dust processing has occurred in the disk.  The 2015 -- 2016 multi-band lightcurve includes variations in spot coverage, extinction, and accretion, each of which result in different traces in color-magnitude diagrams. This lightcurve is initially dominated by a month-long extinction event and return to the unocculted brightness.  The subsequent light-curve then features spot modulation with a 7.03 day period, punctuated by brief, randomly-spaced extinction events.  The accretion rate measured from $U$-band photometry ranges from $1.3\times10^{-8}$ to $1.1\times10^{-10}$ M$_\odot$ yr$^{-1}$ (excluding the highest and lowest 5\% of high- and low- accretion rate outliers), with an average of $4.7 \times 10^{-9}$ M$_\odot$ yr$^{-1}$.   A total of 50\% of the mass is accreted during bursts of $>12.8\times10^{-9}$ M$_\odot$ yr${^{-1}}$, which indicates limitations on analyses of disk evolution using single-epoch accretion rates.
\end{abstract}

\shortauthors{Guo et al.}

\section{Introduction}

Classical T Tauri stars (CTTSs) are low mass young stars surrounded by an accretion disk.  The stellar magnetic field truncates the disk at a few stellar radii and channels gas from the disk onto the star \citep[e.g.][]{Camenzind1990,Koenigl1991,Shu1994}.  The measured strengths and geometries of magnetic fields and the profiles of emission and absorption lines are consistent with expectations of the magnetospheric accretion model \citep[e.g.][]{Johns-krull2007,Donati2009,Hartmann2016}.  Magnetohydrodynamic (MHD) simulations of magnetospheric accretion suggest that the accretion flow may be stable or unstable, depending on the accretion rate, the magnetic field strength and morphology, and the inclination angle between stellar spin and magnetic dipole \citep[e.g.][]{Romanova2013,Blinova16}. 

Photometric variability of T Tauri stars has been studied for decades \citep{Wenzel1969, Grinin1988, Herbst1994, Bouvier2013, Cody2017}. When star-disk interactions are steady, an accretion column and the associated inner disk warp rotates around the star, periodically occulting the central star \citep[e.g.][]{Bouvier2007,McGinnis2015}. In non-steady accretion, these extinction events may appear more stochastically and last for days, months or even years.  The obscure dust is located in a persistent puffed-up disk and inner-rim \citep{Dullemond2003, Ke2012}, a warp induced by binarity \citep{Hamilton2001}, a disk instability at larger distances \citep{Zhang2015}, or perhaps even a non-axisymmetric bridge that links an inner disk with an outer disk \citep{Loomis2017}.  The changes in the height of the inner disk has also been seen in anti-correlated variability of near- and mid-IR disk emission \citep{Espaillat2011}, with a possible relationship to accretion rate \citep{Ingleby2015}.
The disk interpretation is challenged in one case (J1604-2130) by the measurement of a face-on inclination of an outer disk \citep{Ansdell2016B}.  In a second case (RW Aur), the occultation source is uncertain and may be a dusty wind \citep{Petrov2015,Schneider15rw}, a tidal encounter of the secondary star \citep{Dai2015}, the combination of occultation and time-variable accretion \citep{Takami2016}, or partial occultation of the inner disk \citep{Facchini2016}. 

In this paper, we focus on short- and long-term extinction events detected in one CTTS, GI Tau.
Stars with short-duration (1--5 d) extinction events, called {\it dippers}, are obscured by dust structures at or near the disk truncation radius \citep[e.g.][]{Alencar2010, Cody2014, Scaringi2016}. 
AA Tau is the historical prototype for dippers \citep[e.g.][]{Bouvier1999, Bouvier2003}.  Periodic and quasi-periodic dippers have a periodicity distribution consistent with the distributions of stellar rotations \citep{Cody2014}.
 Long-term extinction events, called {\it faders}, occur when the star is occulted by disk components for weeks-to-years \citep[e.g.][]{Bouvier2013, Findeisen2013, Rodriguez2015, Rodriguez2016, Loomis2017}; KH 15D is the prototype for faders \citep{Hamilton2001}. Some stars, including AA Tau, have exhibited both types of extinction events.  Deep extinction events have also been called Type III variables or UXors \citep{Herbst1994}, especially when the occulted object is a Herbig AeBe star \citep[e.g.][]{Grinin1994UX,Natta1997}.

\begin{figure}[!t]
\includegraphics[width=3.in,angle=0]{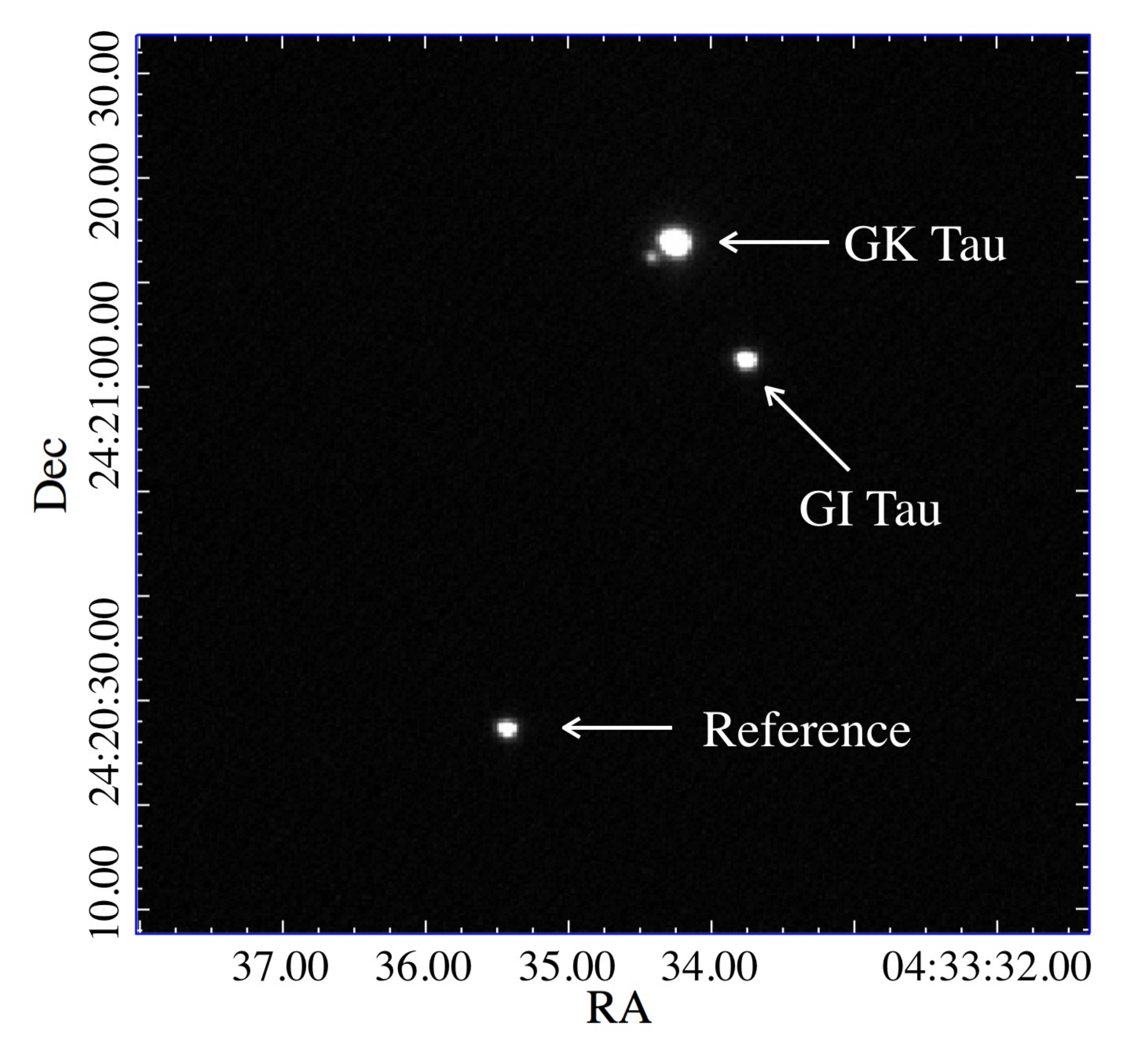}
\caption{$V$-band image of GI Tau and GK Tau obtained using SNIFS at the UH88 telescope.  GI Tau, GK Tau and its close visual companion, and one of the reference stars are marked in the image.}
\label{fig:field} 
\vspace{0cm}
\end{figure}

In the midst of this extinction variability, emission is also always changing because of unstable accretion and spot rotation.  Accretion variability is common on young stellar objects, as 10\% of CTTSs have similar bursty lightcurves \citep{Findeisen2013, Cody2014, Stauffer2014, Cody2017}. The variable accretion process appears as changes in excess continuum and line emission above the photosphere \citep[e.g.][]{Alencar2012,Fang2013,Costigan2014} and the corresponding changes in photometry \citep{Venuti2014,sousa2016,Stauffer2016, Tofflemire2017}, driven by either unsteady star-disk connections \citep[e.g.][]{Romanova2013} or changes in the disk density at the inner rim \citep{Robinson2017}.  Spot modulation is also commonly seen among young stars with typical variations of $\Delta V \lesssim 0.5$ mag \citep[e.g.][]{Herbst1994,Grankin2007},  although spots in lightcurves of some CTTSs can be difficult to distinguish from extinction and accretion variations.
Extinction, accretion, and spot variability each have particular patterns in high-time resolution photometry \citep{Alencar2010,Morales2011, Alencar2012, Cody2017}, multi-band photometry \citep{Herbst1994,Grankin2007,Venuti2015}, and spectroscopic monitoring \citep{Bouvier2007}. 

In this paper, we describe and analyze multi-band optical monitoring of the CTTS GI Tau obtained over two years. Our work provides a method to identify the variation mechanisms by the color information and probe the star-disk interaction at the inner edge of circumstellar disk. The paper is organized as follows. In Section 2, we describe our observation and data reduction. The photometric results and periodicity analysis are described in Section 3. In Section 4, we analyze this photometric variability in terms of the warp size and changes in accretion.

\begin{table*}
\centering
\caption{Summary of Observations}
\renewcommand{\arraystretch}{1.3}
\label{tab:observation}

\begin{tabular}{lcccccccccc}
\hline
\hline
  Telescope & Location     & Diameter (m) & Pix size ($^{\prime\prime}$) & Field of View & No. Ref & Filter & Nights of Obs.      & No. Visits / Night \\
  \hline
   2014 -- 2015  & &  &  &    &  & &\\
   \hline
    UH88 & Maunakea, Hawaii & 2.2 & $0.27$ &  $9.3^{\prime} \times 9.3^{\prime}$ & 2  & $V$ &18    & 1 - 6 \\  
  YNAO  & YNAO, China  & 1 & $0.41$ &  $7.3^{\prime} \times 7.3^{\prime}$ & 2 & $V$ &4  & 4\\
  AZT-11 & CrAO, Russia &1.25  &  $0.62$&  $10.6^{\prime} \times 10.7^{\prime}$ & 3 & $V$ &6   & 1\\
  OAN-SPM (0.84)  & SPM, Mexico & 0.84   &$0.44$&  $7.6^{\prime} \times 7.6^{\prime}$ & 4  &  $V R$  & 4   & 2 hrs$^*$\\   
  HCT  & Hanle, India  & 2  & $0.30$ &  $10.2^{\prime} \times 10.2^{\prime}$& 4 &  $V$ &23  & 1\\
    \hline
  2015 -- 2016& & & & &  & & &\\
\hline
  OAN-SPM (1.5) & SPM, Mexico & 1.5  &$0.32$ &  $5.4^{\prime} \times 5.4^{\prime}$ & 1   & $V I$ & 34 & 2 hrs$^*$\\      
  HCT  & Hanle, India &2   &$0.30$ & $10.2^{\prime} \times  10.2^{\prime}$ & 4 & $U V I$ & 23   & 1 - 3\\
   SLT  & Lulin, Taiwan &0.40  &$0.78$ &  $26.8^{\prime} \times  26.8^{\prime}$ & 4 & $U B V I$ & 74  & 1 - 3 \\
  NOWT  & XAO, China &1   &$1.13$& $1.3^{\circ} \times 1.3^{\circ}$ & 4  &  $B V R I$ &5  & $>$ 5 hrs$^*$\\
  JCBT  & VBO, India & 1.3   &$0.24$& $16.5^{\prime} \times 8.6^{\prime}\,\,$ & 4  &  $B V I$ & 20  & 1\\
  TST  & CTIO, Chile & 0.6  &$0.63$ &  $22^{\prime} \times 22^{\prime}$ & 4 &  $B V R I$ & 45  & 1\\
   NBT & Xinglong, China & 0.85  &$0.91$  & $30^{\prime} \times 30^{\prime}$& 4  &  $U B V R I$ &10   & $>$ 5 hrs$^*$\\  
  TNO & TNO, Thailand & 0.5   &$0.63$ & $21.5^{\prime} \times 21.5^{\prime}$ & 4 &  $B V I$ & 21  & 1 - 2\\
  \hline
  \hline
\end{tabular}
\begin{flushleft}
  \footnotesize {\bf UH88}:  University of Hawaii 2.2 meter telescope. {\bf YNAO}: 1 m RCC-telescope at Yunnan Astronomical Observatory, Kunming, China.{\bf AZT-11}: 1.25 m telescope at Crimean Astronomical Observatory, Russia. {\bf OAN-SPM}: 0.84 m and 1.5 m telescope at Observatorio Astronomico Nacional, Sierra San Pedro M\'artir, Mexico. {\bf HCT}: 2 m Himalayan Chandra Telescope at Indian Astronomical Observatory, Hanle(Ladakh), India. {\bf SLT}: 40 cm telescope at Lulin Observatory, Taiwan. {\bf NOWT}: Nanshan One meter Wide-field Telescope at Xinjiang Astronomical Observatory, Urumqi, China. {\bf JCBT}: 1.3 m J.C. Bhattacharya Telescope at  Vainu Bappu Observatory, Kavalur, India. {\bf TST}: 0.6 m Thai Southern Hemisphere Telescope (PROMPT-8), operated by the Skynet Robotic Telescope Network, at the Cerro Tololo Inter-American Observatory, Chile. {\bf NBT}: 85 cm reflection telescope at Xinglong Station of the National Astronomical Observatories of China. {\bf TNO}: 0.5 m telescope at Thai National Observatory, National Astronomical Research Institute of Thailand (NARIT). \\
  $*$: represents consecutive observation for X hours. 
  \end{flushleft}
\end{table*}

\section{Observations}

\subsection{Properties of GI Tau}

GI Tau is a Classical T Tauri star associated with the B18 cloud in Taurus star forming region \citep{Myers1982,Kenyon2008} and is separated by 13 arcsec from a wide companion, GK Tau (Figure \ref{fig:field}; see, also, e.g., \citealt{Kraus2009}).  GI Tau has a circumstellar disk \citep[e.g.][]{Kenyon1995,Luhman2010,Rebull2010} and ongoing accretion \citep[e.g.][]{Valenti1993,Gullbring1998}. The average VLBI parallax distance of 140 pc to the Taurus star-forming region \citep{Loinard2007, Torres2009, Torres2012} is adopted for the distance to GI Tau.

Companion searches with high resolution near-IR imaging \citep[e.g.][]{Daemgen2015} and high-resolution spectroscopy \citep{Nguyen2012} have yielded non-detections, indicating that GI Tau is likely a single star.  A $\sim 7$~day period has been detected in some epochs \citep{Vrba1986,Herbst1994} but is absent in other epochs \citep[e.g.][]{Grankin2007,Rodriguez2017KELT}, perhaps because spot changes may be masked by complications in the lightcurve from extinction and accretion variability.

The estimated spectral type of GI Tau ranges from K5 -- M0.5 \citep{Rydgren1976, Herbig1977, Cohen1979, Hartigan1994, taguchi09, Herczeg2014}, with differences caused by methodology and a non-uniform temperature distribution on the stellar surface \citep[see, e.g.,][]{Gully2017}.   Extinction events have been previously detected from photometry \citep{Herbst1994, Grankin2007, Rodriguez2016KELT}. In three optical spectra, \citet{Herczeg2014} found that fixing the spectral type to a single value required an extinction that varied from $A_V=1.05$ to $2.55$ mag.  Our analysis in \S 4.3 indicates a minimum $A_V=0.75-1.0$ mag, which is likely interstellar; any additional extinction is likely caused by the disk.

Adopting a spectral type of M0.4 ($T_{\rm eff} = 3828$ K) and $\log (L/ \rm L_{\odot})=-0.25$ (\citealt{Herczeg2014}; see also \citealt{Grankin2016}), the mass and age are 0.53 $\rm$ M$_{\odot}$ and 1.4 Myr as inferred from the pre-main sequence evolutionary tracks of \citet{Baraffe2015}, and 0.92 M$_{\odot}$ and 4 Myr from the magnetic tracks of \citet{Feiden2016}. 
These parameters are sensitive to the unknown spot properties of the star \citep{Gully2017}.   However, dynamical masses measured from disk rotation around stars of similar spectral types lead to masses of 0.60 -- 0.95 M$_\odot$ \citep{Simon2017}. 

 The disk inclination has not been measured.  Given a radius $R=1.7$ R$_\odot$, rotational period $P_{rot} =7.03 \pm 0.02$ d (see \S 3.1), and stellar rotational velocity $v \sin i=12.7 \pm 1.9$ km s$^{-1}$ \citep{Nguyen2009}, the stellar inclination is $>60^{\circ}$ \citep[see also][]{Johnskrull2001}.  Broad redshifted absorption in \ion{He}{1} $\lambda10830$ has a similar profile as that seen in AA Tau \citep{fischer08} and supports this high inclination.

\subsection{SNIFS Photometry and Spectroscopy}

We obtained spectra and photometry of GI Tau with the Super-Nova Integral Field Spectrograph \citep[SNIFS][]{Aldering2002, Lantz2004} from 26 Nov. to 15 Dec. 2014.  SNIFS is an Integral Field Spectrograph  on the UH 88-inch telescope on Maunakea that produces $R\sim1000$ spectra from 3200 to 10000 \AA\ over a $6^{\prime\prime} \times 6^{\prime\prime}$ field-of-view (FOV).  Short acquisition images were obtained with a $9.6^{\prime} \times 9.6^{\prime}$  FOV imager with $V$-band filter and are used here for photometry.

The full set of our SNIFS observations include spectroscopic monitoring of $\sim 30$ CTTSs.  GI Tau was initially selected as a target based on past identification of extinction events \citep[see, e.g.][]{Grankin2007,Herczeg2014}.  We detected a deep extinction event at the beginning of our SNIFS campaign and decided to intensively monitor GI Tau for the remainder of our campaign.    {Two spectra from this spectroscopic monitoring campaign are analyzed in this paper (see \S 2.5).}

\subsection{Subsequent photometric campaigns (2014 -- 2016)}

Following our SNIFS photometry, we monitored GI Tau from 2014 -- 2016 with eleven other telescopes.  The details of the telescopes, instruments, and observations are described in Table \ref{tab:observation}.  The complete set of photometry is listed in an online Table.

From 16 Dec. 2014 (MJD 57007) until 25 Mar. 2015 (MJD 57108), photometry was obtained in the $V$-band filter with a cadence of 1 -- 2 visits per night. From Oct. 2015 -- Feb. 2016, multi-band photometry was obtained in $B$, $V$, $R$, and $I$ bands, and $U$ when available.  Different observational strategies were set based on the time allowance of each telescope. SLT, 1 m Thailand Southern Telescope, and 1.3 m JCBT observed the selected field 1 to 3 times on each clear night. The 0.5 m at TNO and 2 m HCT also contributed weeks-long observations.  The NOWT \citep{Liu2014} and NBT monitored GI Tau for 4--6  hrs for 7 and 3 consecutive nights, respectively, to measure variations on short timescales.

\subsection{Data Reduction of Photometry}

The data were reduced with custom-written routines in IDL.  The images were corrected for detector bias, flat-field, and cosmic rays.  The stellar brightness of GI Tau, GK Tau, and many field stars in the frame are measured with aperture photometry.   For field stars, the sky background is measured in an annulus with 8 arcsec inner radius and 10 arcsec outer radius around the star.  Since the distance between GI Tau and GK Tau is only 13.2 arcsec, the background levels are adopted directly from the sky background of the nearby reference star.  The counts for each star are then extracted using a radius equal to two times the seeing (in FWHM), with an upper limit on radius of $6\farcs5$ arcsec.  Photometry with fixed apertures of $1$, $3$, and $6^{\prime\prime}$ and PSF-fitting yield results that are generally consistent with our approach, but with larger standard deviations in the photometry.

\begin{figure}[!t]
\centering
\includegraphics[height=3.in,angle=0]{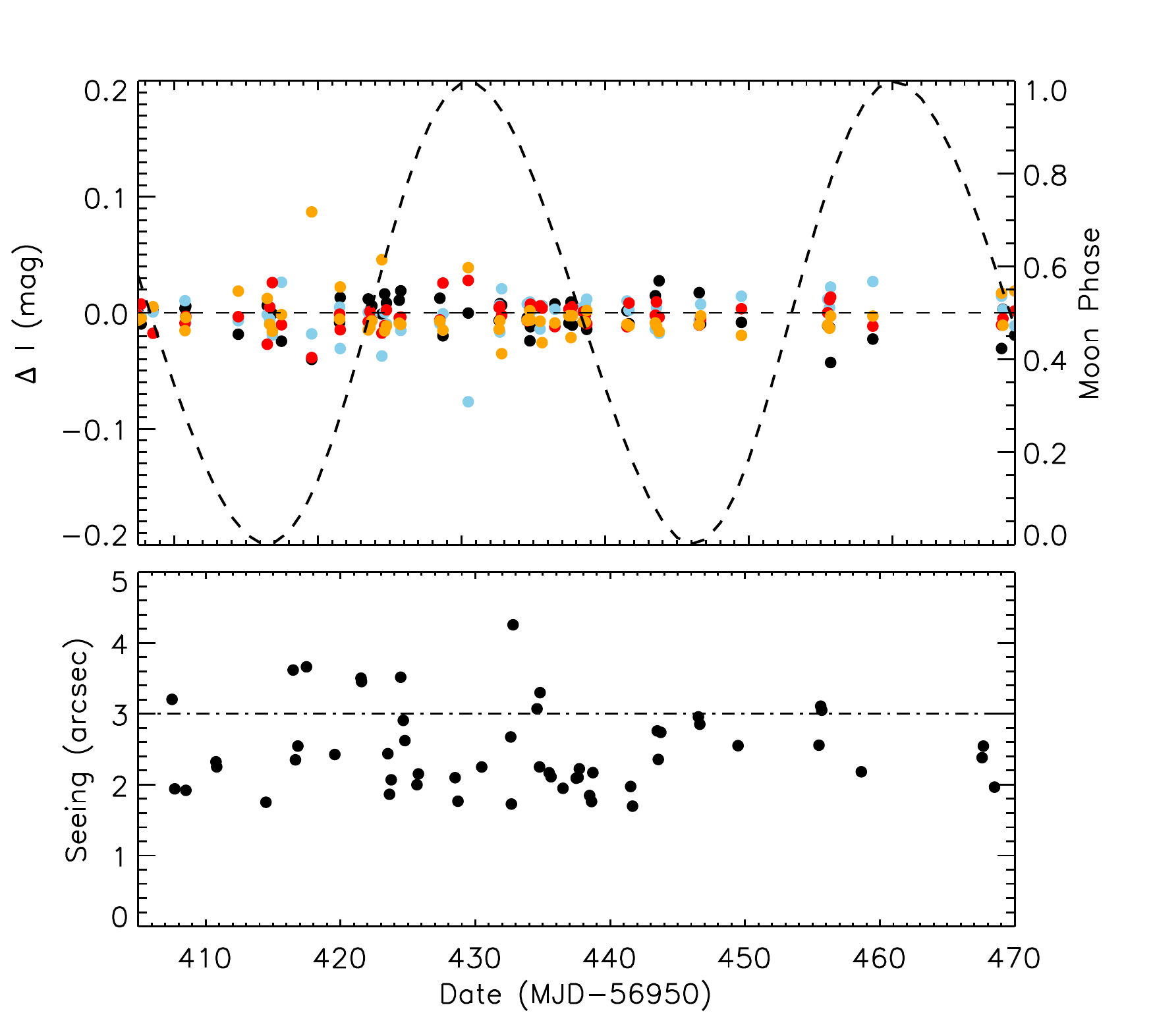}
\caption{\emph{Top:} The accuracy of the $I$-band photometric calibration of the four reference stars (separated by different colors) taken by SLT, plotted as the difference between each observation and the median magnitude, $\Delta I$. The standard deviations of each reference stars are 0.016, 0.018, 0.013, and 0.018 mag. The lunar phase is shown by a dashed black curve. \emph{Bottom:} The seeing during each observation, with a horizontal dot-dashed line indicating 3$^{\prime\prime}$.}
\label{fig:ref} 
\end{figure}

Four bright stars are identified as non-variables (Figure \ref{fig:ref}) and are selected as reference stars to calibrate the $BVRI$ photometry of GI Tau. The measured standard deviations of all reference stars are 0.017 mag in $I$, 0.028 mag in $V$ and 0.042 mag in $B$-band, after excluding the images obtained during the full moon. The measurements are less reliable ($\Delta I > 0.05$ mag) in observations with seeing larger than 4$^{\prime\prime}$.  The number of reference stars used for each telescope depends on the FOV and is listed in Table \ref{tab:observation}.

In $U$-band observations, only one field star, with $m_U$=13.50 mag\footnote{This $U$ band measurement was measured by \citet{Audard2007} with the XMM-Newton Optical and UV Monitor ($U_{\rm OM}$).  With a spectral type of B8, the offset between $U_{\rm OM}$ and Johnson U system of $U - U_{\rm OM} \sim -0.02$ is small and is ignored in our analysis.}, that is located within the $10^{\prime} \times 10^{\prime}$ FOV is bright enough to be used as a calibrator.  Unsaturated images in $B$,$V$ and $I$-band indicate that this calibrator is not variable. The accuracy of our $U$-band observations is typically limited to $\sim 0.05$ mag by the S/N of GI Tau.  The differential effects of telluric absorption versus airmass are not corrected.

A reflection nebulosity around GI Tau and GK Tau \citep{Arce2006} is detected in stacked images, with a surface brightness of $I=22.8$ mag/arcsec$^2$ and $B=25.5$ mag/arcsec$^2$.  The flux contribution from the nebulosity within a $6\farcs5$ radius aperture is $17.5$ mag in the $I$-band and $20.2$ mag in $B$-band, or $\sim 4$ mag fainter than the faintest measurements of GI Tau.  Compared with the photometric accuracy and variability of GI Tau, the differential flux contribution from the nebulosity introduced by the use of different aperture sizes is negligible.

For absolute photometric calibration, we observed the GI Tau field and the region PG 02331 from \citet{Landolt1992} at a range of airmasses with the 2 m Himalayan Chandra Telescope on 1 Dec. 2015.  The atmospheric extinction and instrument coefficients are measured from PG 02331 and applied to  bright stars in the GI Tau field. The standard magnitudes of these reference stars are then used to apply the zero-point shifts to each observation obtained by all other telescopes in this study. 

\begin{figure*}[!t]  
\centering
\includegraphics[height=3.in,angle=0]{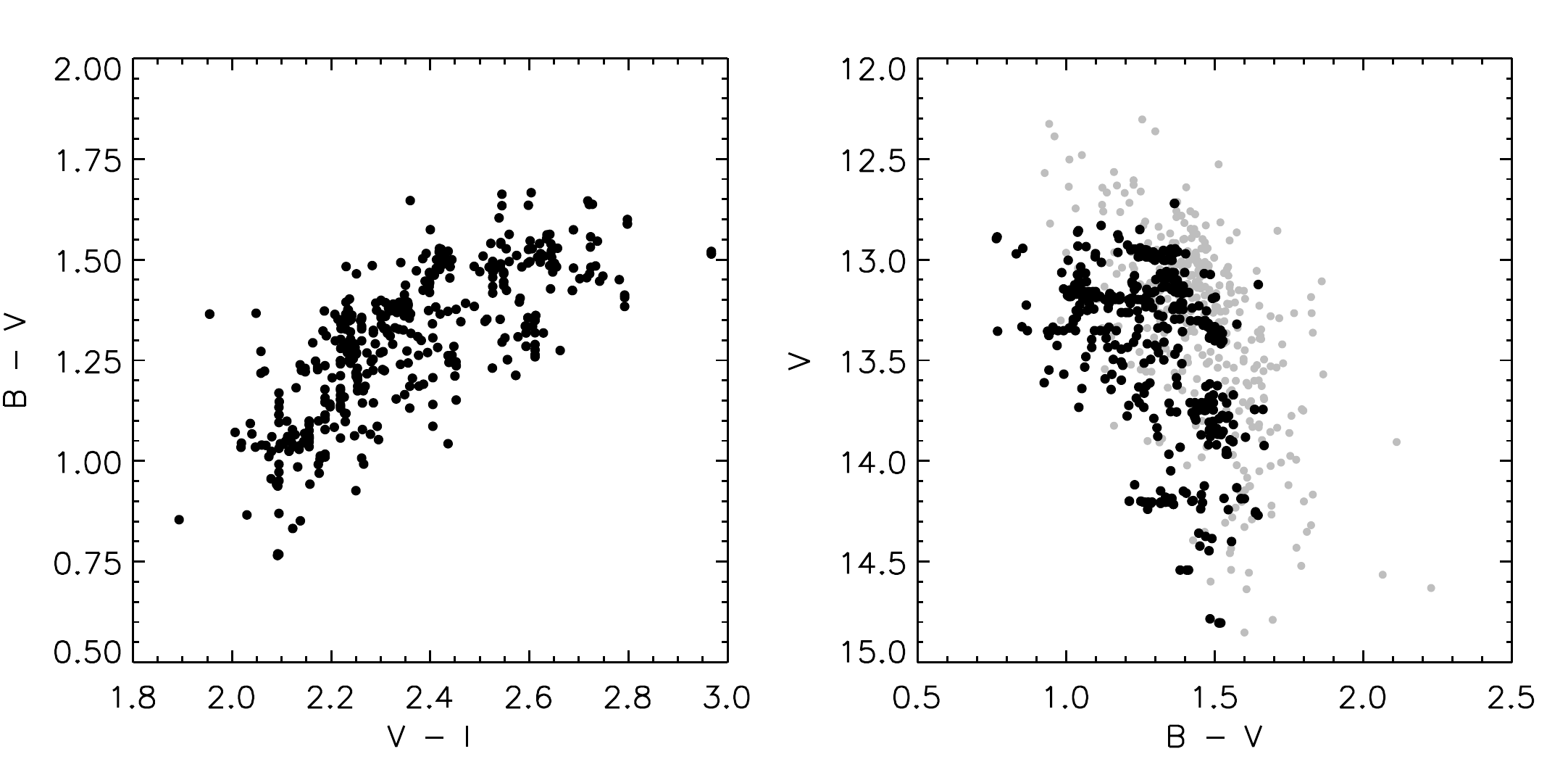}
\caption{Color-color and color-magnitude diagrams of GI Tau during the 2015 -- 2016 observation campaign, with data in our work shown in black dots and archival data from \citet{Grankin2007} shown in grey dots.}
\label{fig:variation} 
\end{figure*}

The absolute photometric calibration accuracy should be $\sim$ 0.02 mag in $V$ and $I$ bands and 0.05 mag in $B$ band, following the uncertainties in the Landolt star calibrations. However, an absolute offset of $0.09$ mag in $V$-band calibration is identified when comparing our photometry to the historical photometry of \citet{Grankin2007} (see~Figure~\ref{fig:variation}) and to the synthetic photometry obtained from our flux-calibrated SNIFS spectra.  The source of this problem could not be identified.  Our relative photometric calibration should be unaffected.  The synthetic $\Delta V$ between our SNIFS spectra is within 0.01 mag of the directly-measured $\Delta V$ obtained in our acquisition images.

\subsection{Data Reduction of Spectroscopy}

The SNIFS spectra of GI Tau and the spectrophotometric standard G191B2B \citep{Oke1990} were reduced with custom-written routines in IDL.  The emission is split at $\sim 5200$ \AA\ by a dichroic into separate red and blue channels.  The raw images consist of 225 separate spectra, each from a given spaxel in the $15\times15$ integral field unit.  The counts in each spectrum are extracted by fitting a cross-spectrum profile, measured from flats, to each wavelength pixel.  The spectra in each spaxel was then wavelength-calibrated to $\sim 10$ km s$^{-1}$ using arc lamps, flat-corrected in each spaxel, and then re-gridded onto the same wavelength scale.

The final spectra are extracted from the data cube by fitting a 2D profile and sky background at each wavelength bin.  The spectra of GI Tau were then flux-calibrated using G191B2B spectra obtained within 1 hr of GI Tau.  The average airmass correction was calculated using spectra of G191B2B over the 20-night run and was then applied to each epoch.  Two spectra were selected for use in this paper because they were obtained in photometric conditions, near in time to the photometric calibrators, and at the local minimum and maximum of the lightcurve.

\begin{figure}[!t]
\includegraphics[width=3.7in,angle=0]{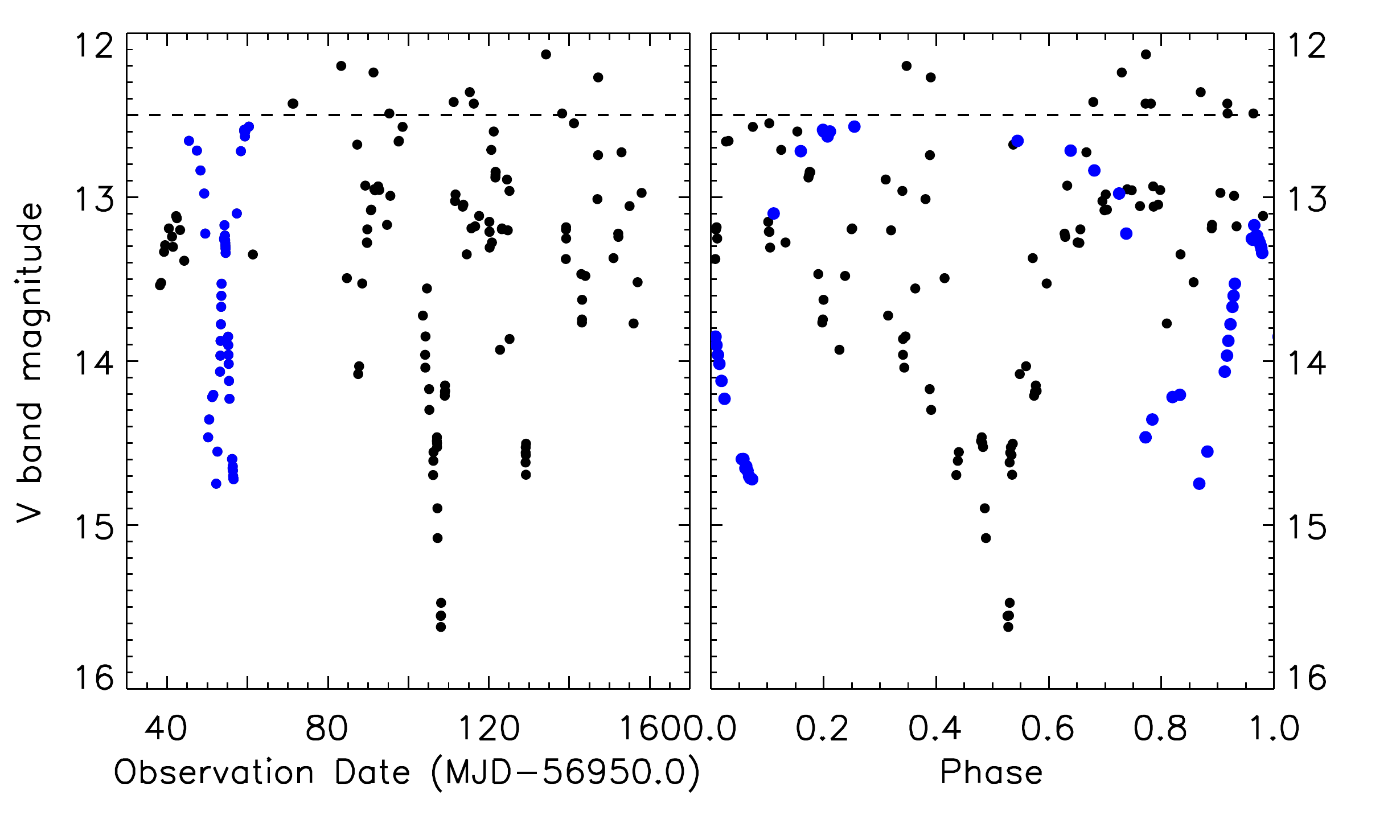}
\caption{$V$-band light curve of GI Tau during 2014 -- 2015 versus time (left) and phase-folded for the    {$\sim 21$} d period (right), and binned to 30 min intervals where relevant.  A `double dip' feature from Day 45 to 61 is shown by blue dots. The horizontal dashed line is the approximate baseline of GI Tau used here to calculate the occultation depth.}
\label{fig:gitau2014} 
\end{figure}

\begin{figure}[!t]
\centering
\includegraphics[width=3.in,angle=0]{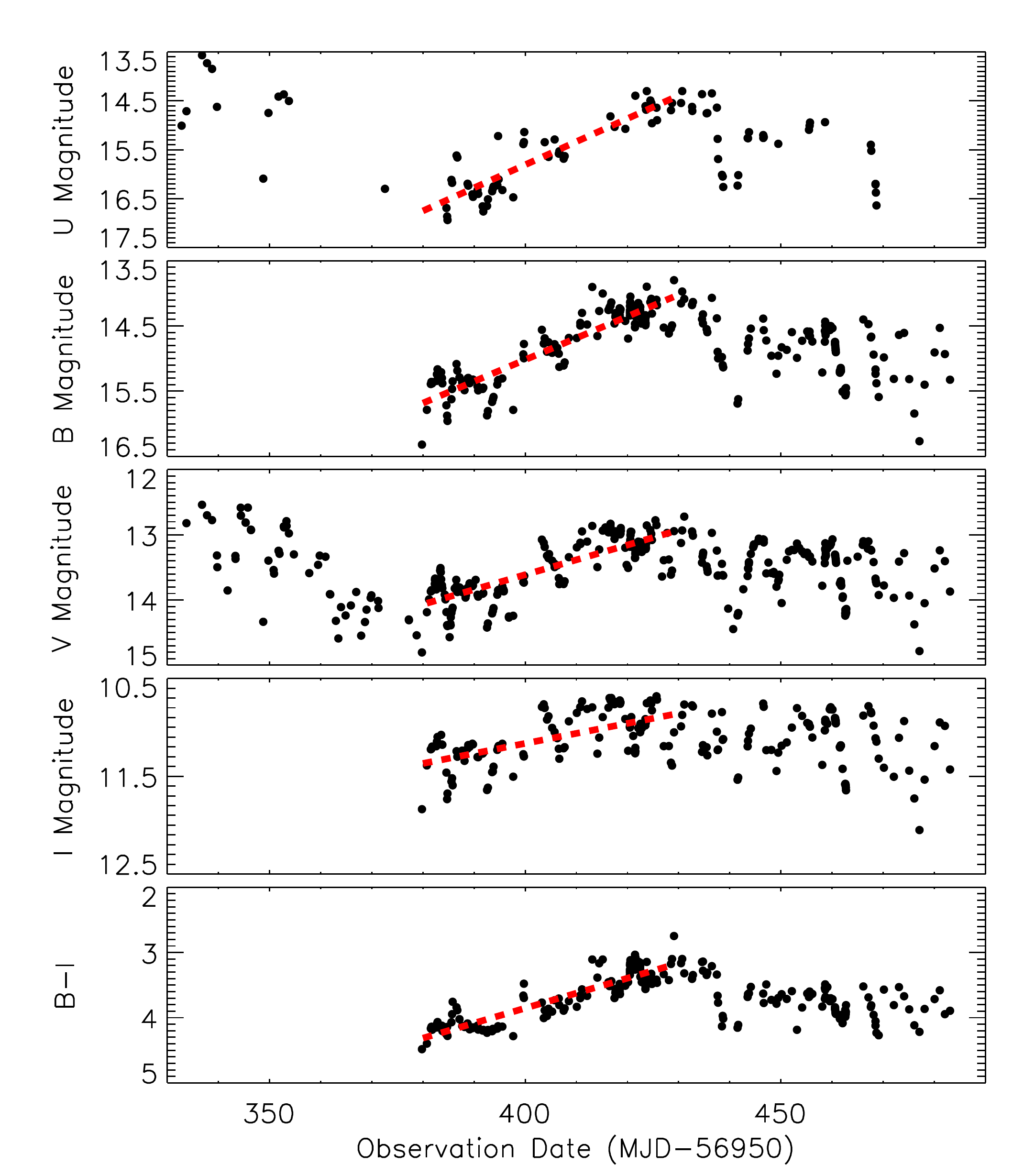}
\caption{From top to bottom, the $U$, $B$, $V$, and $I$-band and $B-I$ lightcurves of GI Tau during the 2015--2016 campaign.  The general brightening that occurred from Day 380 is fit with the red dashed lines.}
\label{fig:gitau2015} 
\end{figure}

\section{Results and Analysis}

In the 2014 -- 2015 light curve of GI Tau, the most prominent features are several extinction events with depths of $\Delta m_{V} > 2.5$ mag and durations of 3 -- 5 days (see Figure \ref{fig:gitau2014}).  The 2015 -- 2016 light curve of GI Tau began with a dim epoch that lasted $\sim 50$ days, followed by a period with smaller periodic brightness variations (Figure \ref{fig:gitau2015}).  

These photometric variations are summarized by the color-color and color-magnitude diagrams in Figure \ref{fig:variation}. The $V$-band brightness varied by 2 mag, the $V-I$ color by 0.8 mag, and the $B-V$ color by 0.5 mag.  The locus of points on the color-magnitude diagram is similar to that seen in long-term monitoring of GI Tau by \citet{Grankin2007}, except for the offset in $V$-band discussed in \S 2.4.

In faint epochs, a `blue turnaround' is seen, in which the color variation is achromatic with further dimming of $V$.  This blue turnaround, also seen in AA Tau \citep{Bouvier1999} and other CTTSs \citep{Grankin2007}, is likely caused by an increased importance in scattered light, since stars with edge-on disks typically appear blue at optical wavelengths \citep[e.g.][]{Padgett1999,Herczeg2014}.  These epochs are ignored when calculating accretion rates.  However, if the bluer colors are caused by higher accretion rates during these faint epochs, then this choice would bias our results.

In this section, we describe how the light curves are combined with the color-color and color-magnitude diagrams are used to identify variability caused by stellar spots, circumstellar extinction events, and accretion bursts.

\begin{table*}
\centering
\caption{Sine fit results}
\renewcommand{\arraystretch}{1.2}
\label{tab:sinfit}
\begin{tabular}{lcccc}
\hline
\hline
  Parameters      & I+poly & I & V & B \\
  \hline
  Period (d)    & 7.03$\pm$0.02 & 7.01$\pm$0.03  & 7.09$\pm$0.08 & 7.20$\pm$0.09 \\
  Frequency (1/d)    & 0.1422$\pm$0.0004 & 0.1426$\pm$0.0006  & 0.140$\pm$0.002 & 0.139$\pm$0.002\\
  Maximum Power: $p_{max}$    & 0.829 & 0.417 & 0.645 & 0.567 \\
  Standard deviation: $\sigma_{p}$ & 0.037 & 0.022 & 0.078 & 0.110\\
  Index: $p_{max}/\sigma_{p}$ &21.82&18.77& 8.26  & 5.15\\ 
  Amplitude (mag)   & 0.24$\pm$0.01 & 0.23$\pm$0.02 &  0.32$\pm$0.09 & 0.41$\pm$0.03\\
  RMS of Residual (mag)    & 0.145 & 0.202 & 0.167 & 0.488\\
  \hline
  \hline
\end{tabular}
\end{table*}

\subsection{Spot modulation in 2015 -- 2016}

\begin{figure*}[!t]
\centering
\includegraphics[width=3.in,angle=0]{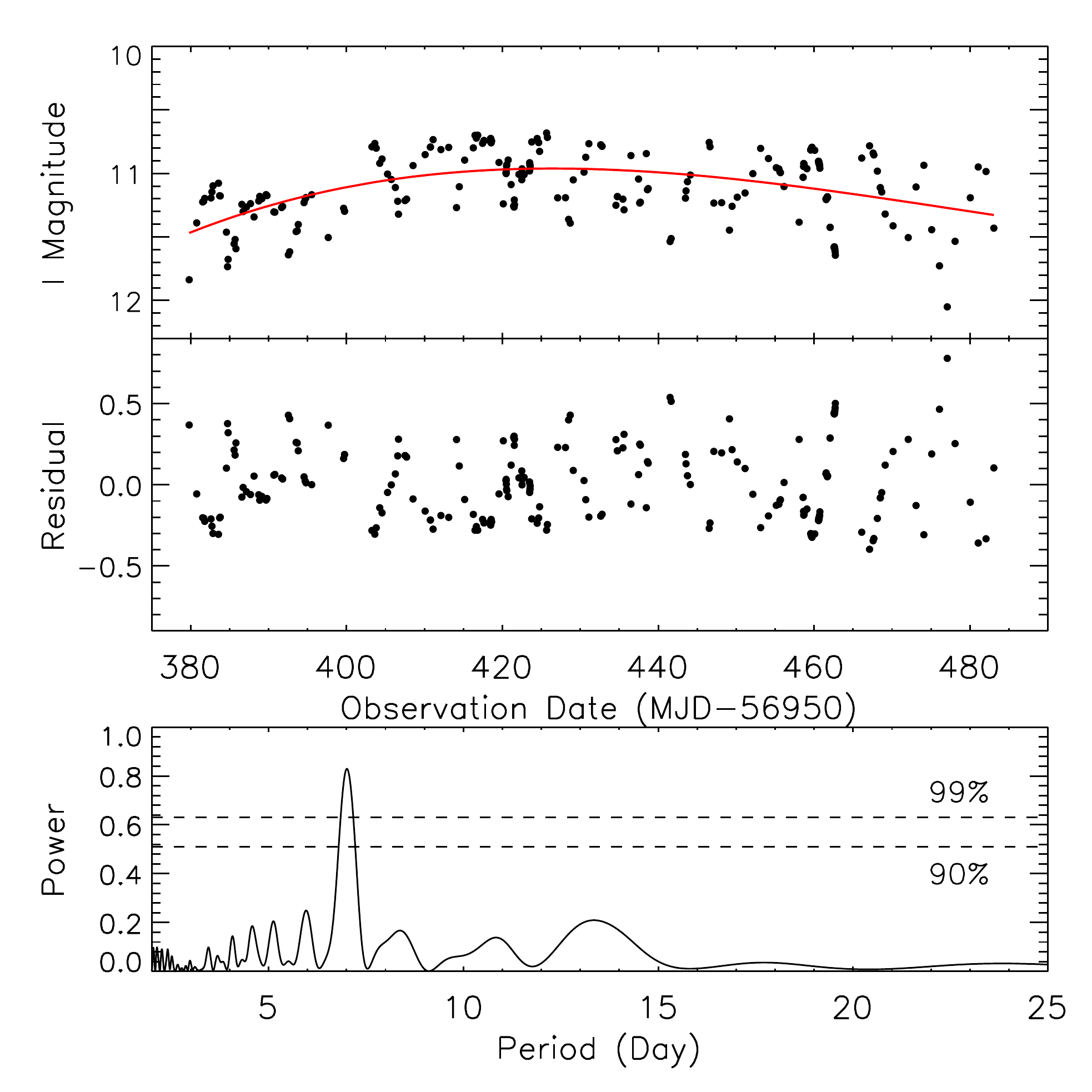}
\includegraphics[width=3.in,angle=0]{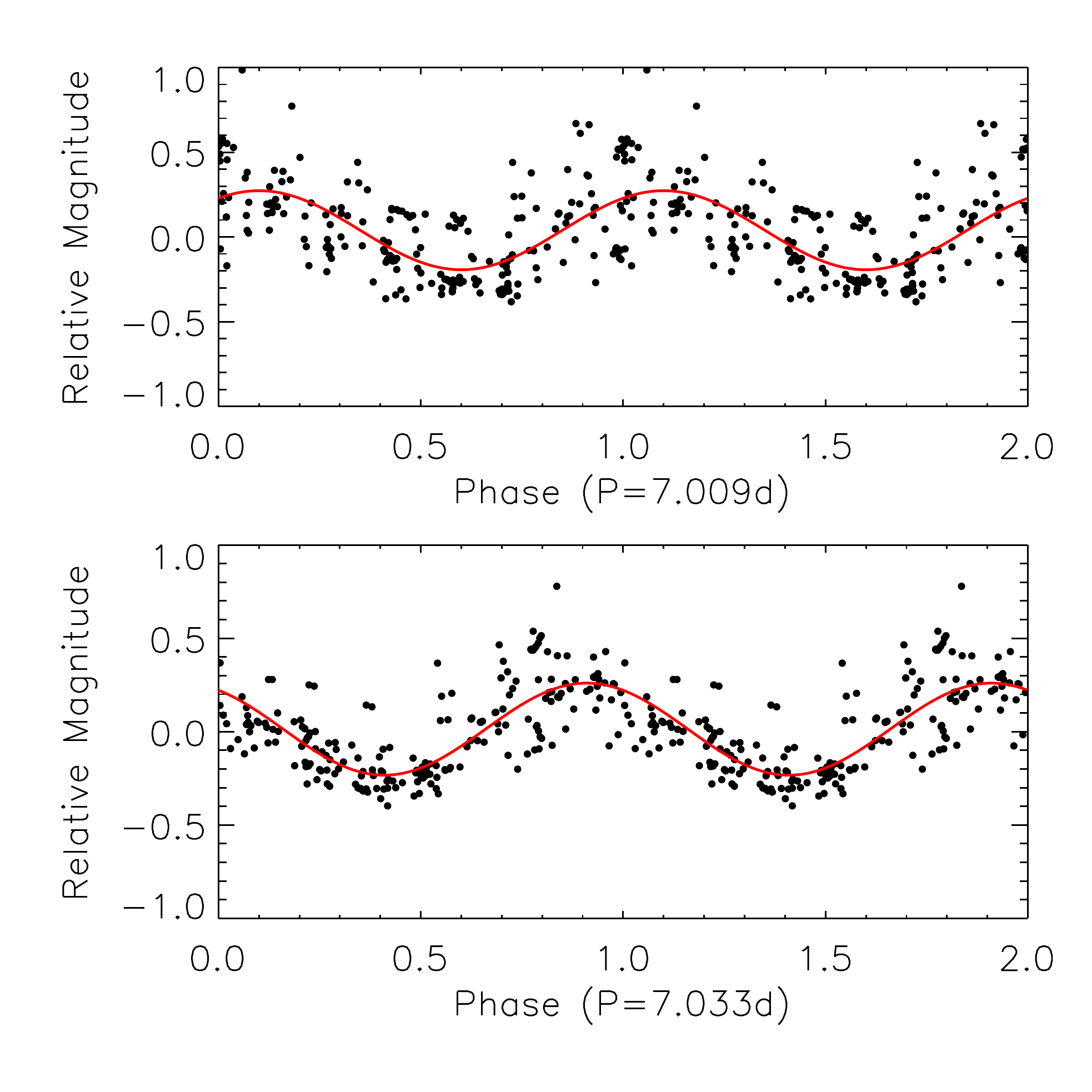}
\caption{{\it Top Left:} The $I$-band lightcurve of GI Tau, with a red line showing a 3rd order polynomial fit to long-term variations. {\it Middle left:} The residual of the fit from in the upper panel. {\it Bottom left:} The periodogram calculated from the light curve in the middle panel, showing a peak at 7.03 d. {\it Top Right:} Phase-folded $I$-band light curve in campaign 2015 -- 2016 using the raw data from the top left panel. {\it Bottom Right:} Phase-folded $I$-band light curve by the residuals from the left middle panel.}
\label{fig:period} 
\end{figure*}

Periodicity in the 2015 -- 2016 lightcurve is most prominent in the $I$-band. The Generalized Lomb-Scargle (GLS) periodogram \citep{Zechmeister2009} of the $I$-band lightcurve yields a best-fit period of $7.03 \pm 0.02$ d, with the error bar adopted from the FWHM of the periodogram profile (Figure~\ref{fig:period}).  Prior to the fit, the long-term trends were approximated as a third-order polynomial and were removed from the data \citep{Zajtseva2010}. Fitting parameters to $B$, $V$, and $I$-band lightcurves are shown in Table~\ref{tab:sinfit}. 

\begin{figure}[!t]
\centering
\includegraphics[width=3.in,angle=0]{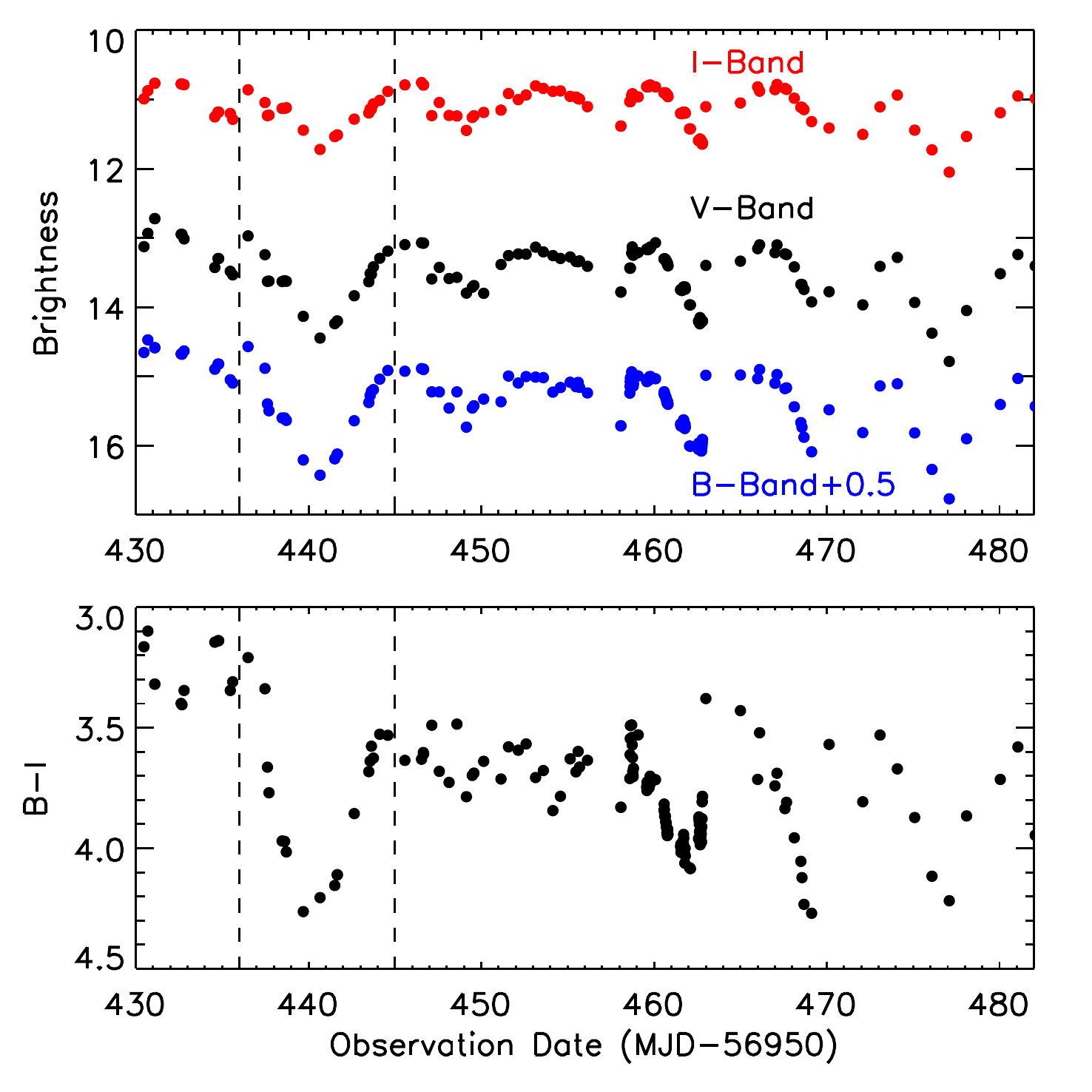}
\caption{\emph{Top}: $B$ (offset by 0.5 mag), $V$, and $I$-band lightcurves of GI Tau between days 430 -- 485, showing a combination of spots and occultation.  {\it Bottom:}  The $B-I$ color, with large dips that indicate occultations.}
\label{fig:gitau3color} 
\end{figure}

 The sinusoidal morphology of the phase-folded light curves indicates the presence of a single large spot, similar to some other young stars with similar spectral types \citep[e.g.][]{Alencar2010,Rebull2016,Gully2017}. The standard deviation of the residual of 0.11 mag is likely caused by extinction and accretion events (discussed in \S 3.2 -- 3.4).  The power of the periodogram, $\zeta = p_{\rm{max}}/\sigma_{p}$, is highest in the $I$-band, since the other bands are more sensitive to accretion and extinction variations.  The variations in the colors are synchronous (Figure~\ref{fig:gitau3color}). 

False-alarm probabilities\footnote{False-alarm probabilities are the fraction of permutations (ie. shuffled time-series) that include a peak higher than that of the periodogram of the unrandomized dataset at any frequency. This therefore represents the probability that, given the frequency search parameters, no periodic component is present in the data with this period. To ensure reliable significance values, the number of permutations was set to 1000. If the false alarm probabilities lie between 0.00 and 0.01, then the quoted period is a correct one with 95\% confidence. The periodogram is computed at 5000 frequencies between 0 and 0.5 d$^{-1}$.} for the period are computed using a Fisher randomization test with input periods between 2 -- 100 days \citep[e.g.][]{Nemec1985}. The 7.03 day period exceeds the 99\% confidence level. This period is consistent with past measurements of the photometric period (Table~\ref{tab:period}).  In other epochs, including our monitoring in 2014 -- 2015 and the 2008 -- 2014 light curves described by \citet{Rodriguez2017}, any modulation from spots is masked by much stronger variability caused by extinction.

\begin{table}
\centering
\caption{Extinction events on GI Tau}
\begin{tabular}{ccccc}
\hline
\hline
Time (MJD-56950) & $V_{\rm min}$ (mag) & $\Delta V$ (mag) &Duration (day) \\
  \hline
  50.2 & 14.34 & 1.84 & 5\\
  56.5 & 14.72 & 2.22 &4\\
  87.5 & 14.07 & 1.57 &$> 3$\\
  108.1 & 15.62 & 3.12 &5\\
  129.2 & 14.70 & 2.20  &-\\
  380.0 & 14.34 & 1.54 & 80 \\
  396.8 & 14.27 & 0.48 & 3 \\
  440.6 & 14.45 & 1.15  & 8 \\
  477.1 & 14.78 & 0.96 & 4 \\
  \hline
  \hline
  \label{tab:dip}
\end{tabular}
\end{table}

\subsection{Extinction events in 2014 -- 2015}

Several photometric dips are shown in the $V$-band light curve of 2014 -- 2015 campaign, with depths of $1.5$ -- $3.1$ mag relative to the out-of-extinction brightness of $\sim 12.5$ mag and durations of  3 -- 5  days (see list of extinction dips in Table~\ref{tab:dip}).

The lightcurve of GI Tau reveals a wide range of durations and frequencies of extinction events. Our initial SNIFS monitoring included a double-dip extinction event, during which the $V$-band emission from the star faded, brightened, and then quickly faded again. The separation of the two minima is 5 days, and the total combined duration of 11 days, longer than one stellar rotation period. The $R_V$ measurement based on spectra will be discussed in \S 4.2.

Subsequent follow-up photometry over the next months led to the detection of four dips with $A_V > 1.5$ mag (see Table \ref{tab:dip}). These dips have a centroid time that repeats with a $\sim 21$ day period.  However, the preceding double-dip is inconsistent with this quasi period.  The extinctions that occur in the following year, described below, are also inconsistent with any periodicity.

\subsection{Extinction events in 2015 -- 2016}

\begin{figure}[!t]
\centering
\includegraphics[width=3.5in,angle=0]{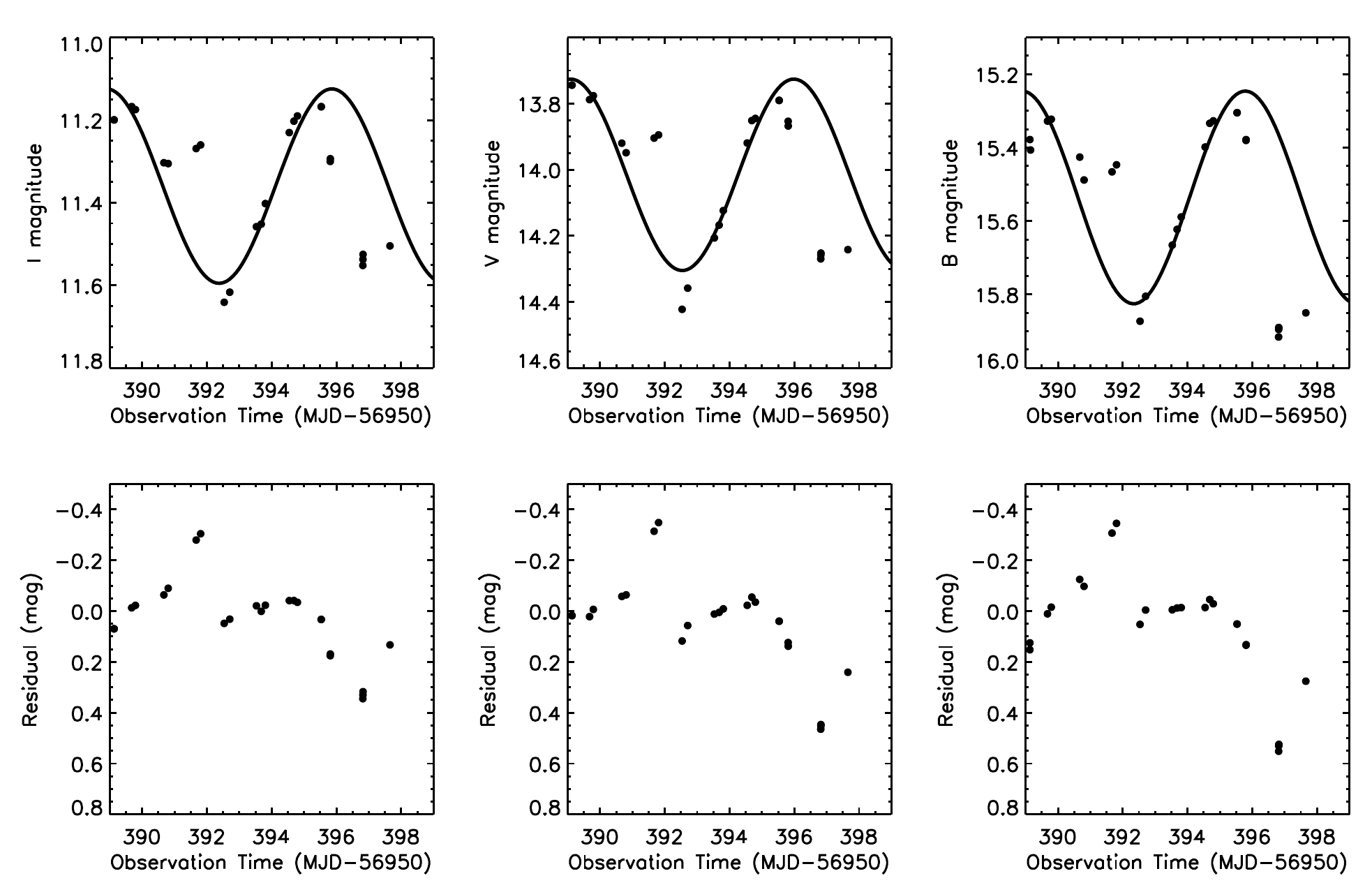}
\includegraphics[width=3.5in,angle=0]{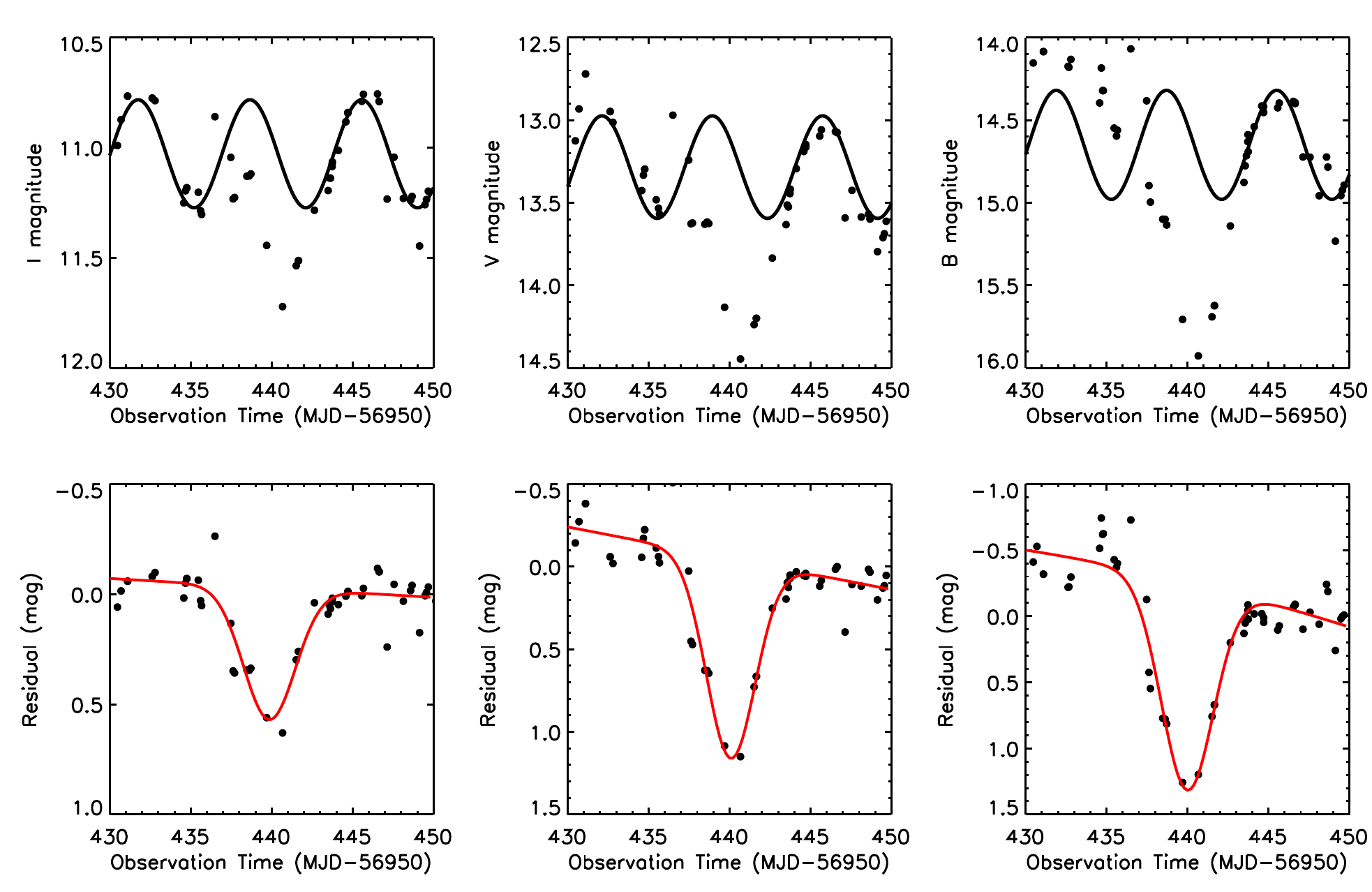}
\caption{{\it Top two panels}: $I$, $V$ and $B$ bands lightcurves of GI Tau from days 391 -- 399, with sinusoidal fits with the 7.02 d period and residuals from the fit. Curves in upper panels show the sine fit as spot modulation. {\it Bottom two panels}: Same as the top set of panels, for days 430 -- 450 and showing a Gaussian profile fit to extinction events in red.}
\label{fig:gi389} 
\end{figure}

The lightcurve during our 2015 -- 2016 campaign is initially dominated by a gradual fade that reaches $\Delta V = 1.5 $ mag and then returns to the bright state, in total covering a period of $\sim 80$ days (Figure~\ref{fig:gitau2015}).  In addition to this months-long fading event, several small and large photometric dips are detected with durations of 3 -- 8 days, after correcting for spot-induced periodicity (see Figures~\ref{fig:gitau2015} and \ref{fig:gi389} and Table~\ref{tab:dip}).

Figure \ref{fig:gitau2015} shows a brief ($\sim 3$ day) dip in the spot-corrected lightcurve at Day 397, with a depth of $ \Delta I = 0.39$ mag, $\Delta V=0.45$ mag, and $\Delta B$=0.56 mag.  A deeper and longer dip occurred around day 440, lasting for $\sim$ 8 days (Figure~\ref{fig:gi389}). Gaussian fits to the dips, as measured after accounting for spot rotation, yield $A_I = 0.60$ mag, $A_V = 1.22$ mag and $A_B = 1.56$ mag and FWHM of $3.73$, $3.52$ and $3.76$ days, respectively. In those fits, the depths are measured relative to the post-dip lightcurve, which is well fit by a sine curve. There is no obvious periodicity of this extinction event.

\begin{figure*}[!htp]
\centering
\includegraphics[width=4.6in,angle=0]{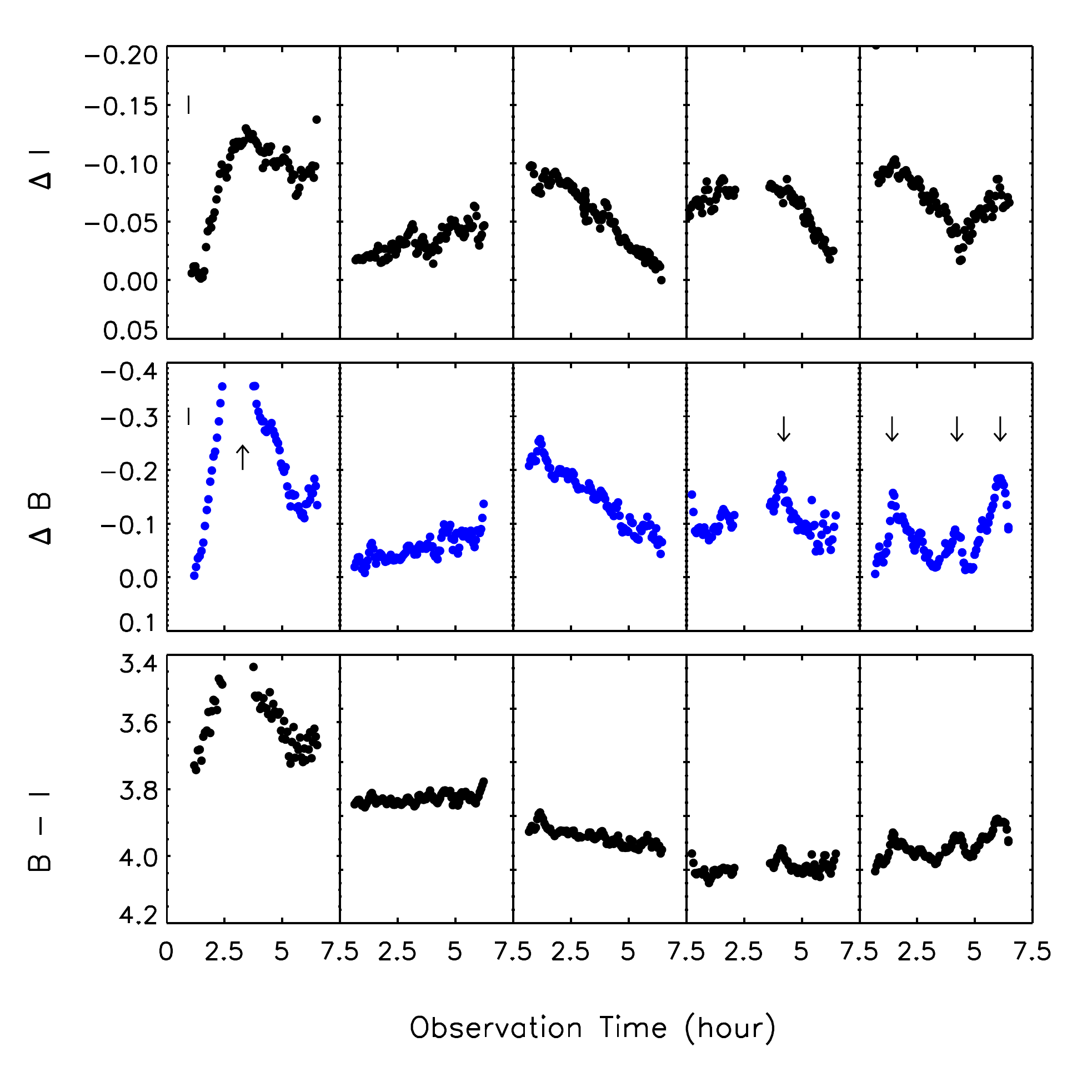}
\caption{$B$, $I$ and $B - I$ light curves of GI Tau from 5 consecutive half-nights using NOWT. The $B$- and $I$-band lightcurves are normalized to the minimum brightness within each day to compare their morphologies. Strong accretion bursts are marked by arrows. Error bars for the $B$ and $I$-band are shown at the upper left corner.} 
\label{fig:accb} 
\end{figure*}

\subsection{Short timescale bursts}

Photometry in the $U$- and $B$-bands is more sensitive to accretion than photometry with longer-wavelength filters.  At short wavelengths, the photospheric emission of GI Tau is faint relative to the continuum emission produced by the accretion shock \citep[see review by][]{Hartmann2016}. In our monitoring, the $U$- and $B$-bands exhibit stronger variations than the $V$ and $I$-bands. 

Our campaign included five nights with constant monitoring of GI Tau on NOWT, during which several short bursts occurred (Figure~\ref{fig:accb}). The largest burst in $B$, detected during the first night, reached a peak $\Delta B\sim 0.3$ mag and lasted $\sim 3.5$ hr.  Four other shorter, smaller bursts are detected in the last two days.   The average duration of these five bursts detected by NOWT is $\sim 1.7$ hr, and the maximum amplitude in $B$-band is 0.31 mag. The change in brightness caused by these accretion bursts are an order of magnitude smaller than those caused by the deep extinctions. The corresponding increases of accretion rate during these bursts are calculated in \S 4.3.  In one case, the $B$-band brightness is consistent with a non-detection, so the minimum and maximum accretion rates before and during the burst are not reported.  These short bursts are attributed here to accretion but could alternately be attributed to stellar flares \citep[e.g.][]{,Kowalski2016, Tofflemire2017, Tofflemire2017b}.

\begin{table*}
\centering
\renewcommand{\arraystretch}{1.2}
\caption{Hours long timescale bursts}
\begin{tabular}{ccccccc}
\hline
\hline
 Time (MJD -- 56950) &  $\Delta B$ (mag)  &  $\Delta V$ (mag) &  $\Delta I$ (mag) & Duration (hour) $ ^{*} $ & $\dot{M}_{\rm acc,min}$$ ^{**} $ &$\dot{M}_{\rm acc,max} $$^{**}$\\ 
  \hline
458.7 &  $>$ 0.31 & 0.32 &	0.12	&2.4 & 2.27 & 7.31\\
461.7 &	0.04	& 0.03 &0.005&	1.0 & $^a$ & $^a$ \\
462.6 &	0.10	& 0.07 &0.02	&2.2 & 8.77 & 11.1 \\
462.7 &	0.06	& 0.02 &-	&1.7 & 11.5 & 14.4 \\
462.8 &	0.17	& 0.10 &0.10&	$>$ 1.44  & 11.6 & 15.4 \\	
  \hline
  \hline
\multicolumn{7}{l}{$*$: Full duration of bursts measured by $\Delta B$ in Figure~\ref{fig:accb}.}\\
\multicolumn{7}{l}{$**$: The mass accretion rates are in unit of $1 \times 10^{-9} \, {\rm M}_{\odot}\, {\rm yr}^{-1}$.}\\
\multicolumn{7}{l}{$^a$: The $B$-band photometry is below the detection limit set in \S 5.3}\\
\label{tab:burst}
\end{tabular}
\end{table*}

\subsection{Color Analysis}

\begin{figure*}[!htp]
\centering
\includegraphics[width=3.3in,angle=0]{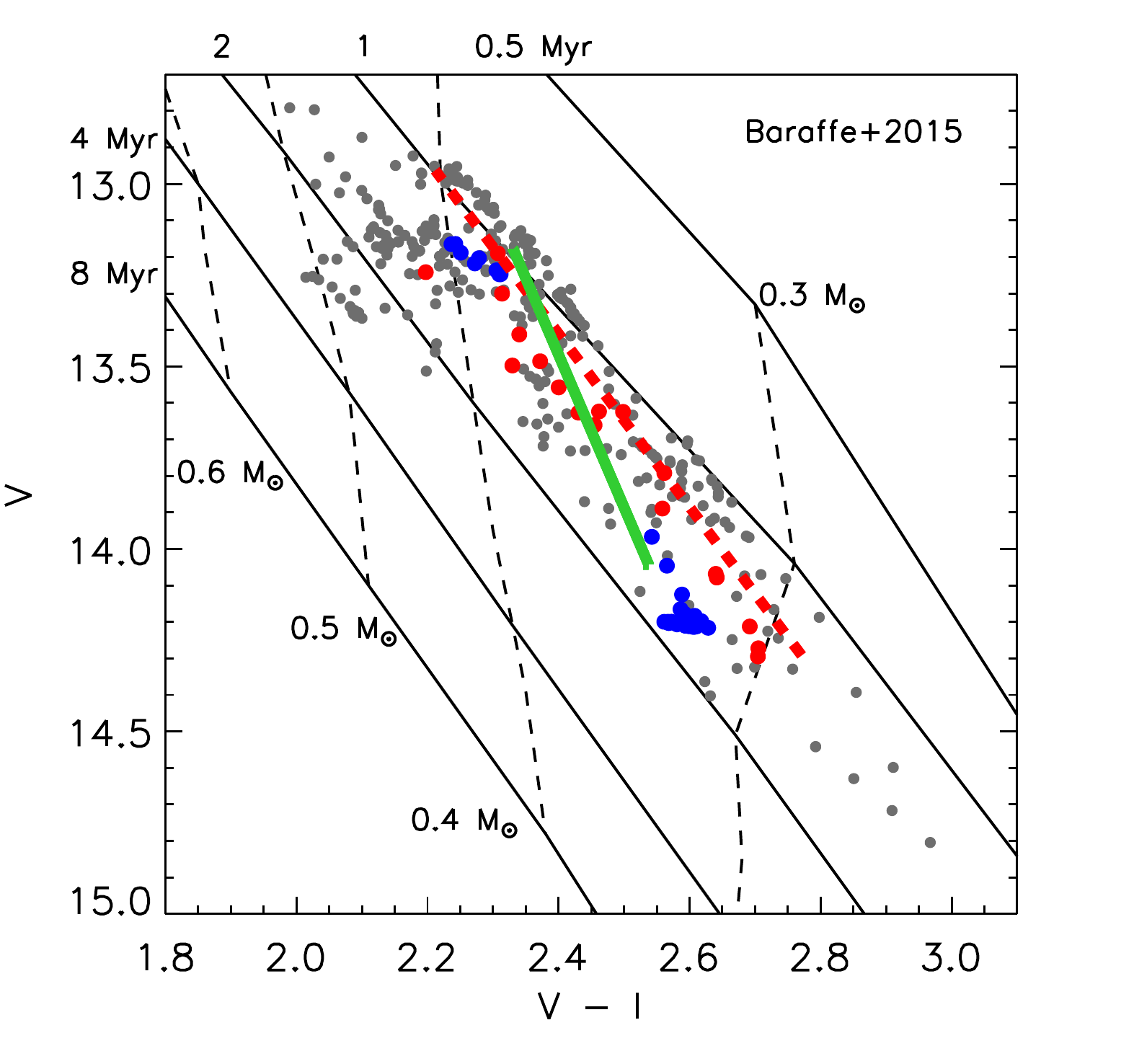}
\includegraphics[width=3.3in,angle=0]{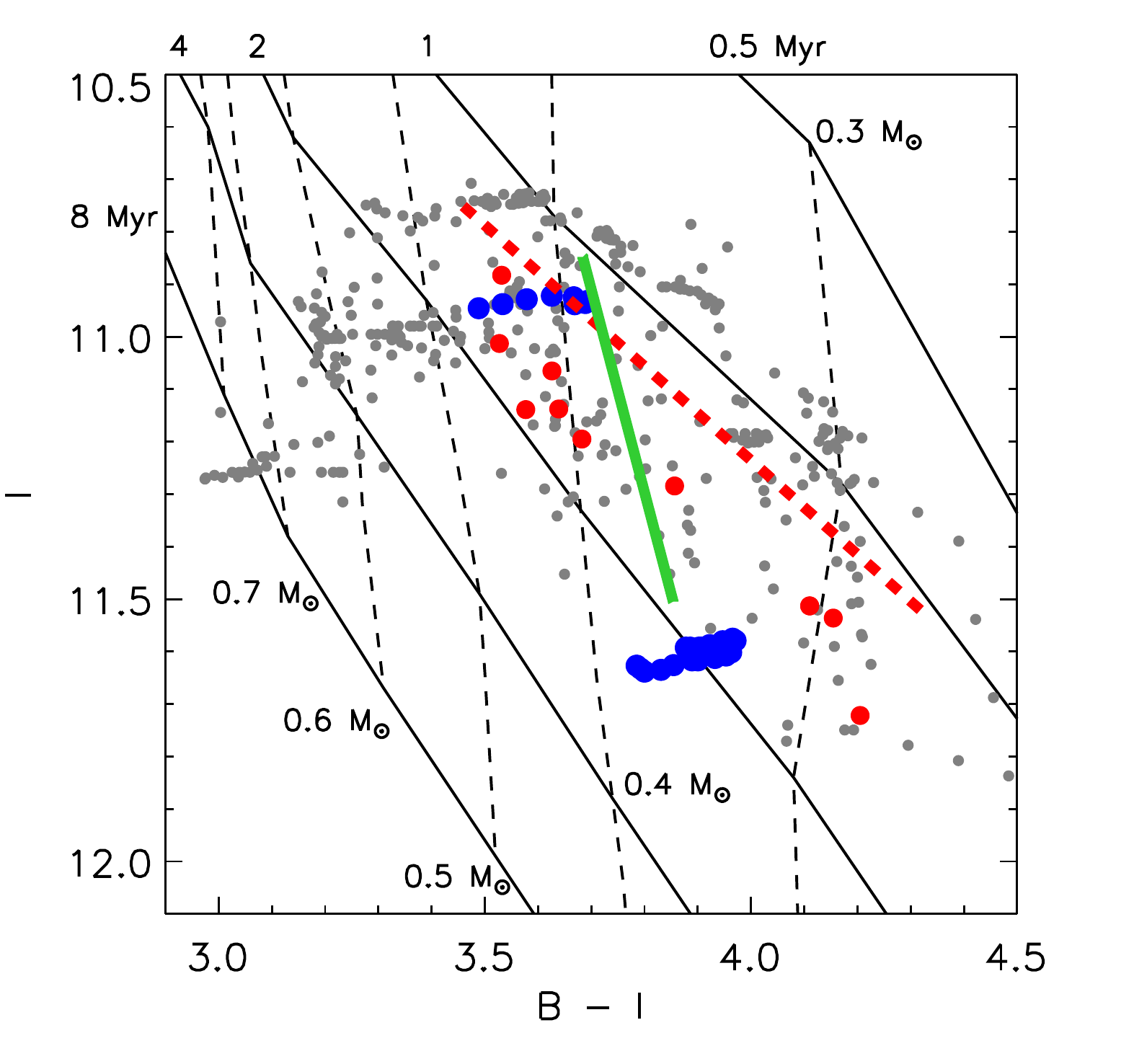}
\caption{{\it Left:} The $V-I$ versus $V$ color-magnitude diagram from our multi-band monitoring of GI Tau, with observed data from 2015--2016 in gray. Pre-main sequence evolutionary models by \citet{Baraffe2015} are presented to show the isochrones and mass tracks shifted to a 140 pc distance. The red dots show the extinction event around day 440. The red dashed line shows the fit to the long time fading event shown in Figure~\ref{fig:gitau2015}. Blue dots are two short accretion bursts detected by NOWT. Spot modulation is shown by the green line.  {\it Right:} The $V-I$ versus $I$ color-magnitude diagram, with the same points as on the left.}
\label{fig:cmd} 
\end{figure*}

Variable extinction, accretion, and spot coverage are all identified from the optical lightcurve of GI Tau. The traces of different phenomena in the color-magnitude diagrams can be used to distinguish the variation mechanisms. In this section, we describe the different signatures that changes in each of these properties imprints in color-color and color-magnitude diagrams (Figure~\ref{fig:cmd}).

The short extinctions dips in the 2015 -- 2016 campaign exhibit similar changes in the color-magnitude diagram  with $\Delta V$ = $2.10 \pm 0.08 \, \Delta (V - I)$ and $\Delta I$ = $0.7 \pm 0.1 \, \Delta (B - I)$.  The long-term variation seen in the first half of the 2015 -- 2016 campaign appears similar to the dips and is also attributed to extinction. These empirical relationships are consistent with expectations for dust reddening. The accretion bursts appear as horizontal changes in $B-I$ versus $I$, indicating that the accretion only has a minor effect on the $I$-band brightness and that the $B-I$ color is a good tracer of accretion. In this case,  accretion is much flatter than extinction in the  $I$ versus $B-I$ diagram (Figure \ref{fig:cmd} and Table~\ref{tab:slope}).  \citet{Venuti2015} obtained similar results in two weeks of monitoring young stars in NGC 2264 with CFHT in $u^\prime$ and $r$ bands.  

As the spot rotates, the $V-I$ colors change by 0.06 mag while the $B-V$ colors change by 0.08 mag. These small color changes during spot modulation are consistent with those of the weak-line T Tauri star LkCa 4 during three decades of photometry \citep{Grankin2008,Gully2017}.  The locus that spot modulation traces on the color-magnitude diagrams has a slope between those of accretion and extinction.  However, since the spot modulation has a unique periodicity, the spot information is readily extracted from a frequency analysis.

Pre-main-sequence stellar evolution tracks from \citet{Baraffe2015} are also presented in the color magnitude diagrams, with colors adopted from \citet{allard14}.  In distant clusters, properties of low mass stars are often inferred from photometry \citep[e.g.][]{Reggiani2011,Jose2016,Beccari2017}.  Extinction events, accretion bursts, and spots each influence the inferred mass and age of member stars.  Extinction curves are parallel to the color isochrone of cool stars in $V-I$ versus $V$ diagram, which indicates that the age determination from $V$ and $I$-band photometry is robust to extinction changes \citep[see also discussion in][]{Sicilia05}.  The age of GI Tau inferred from the \citet{Baraffe2015} models is consistently between 1--2 Myr (see also the age estimation in \S 2.1).  However, the $V-I$ color range introduces uncertainty in mass or T$_{\rm eff}$ estimates when analysis is restricted to photometry,    with larger uncertainties when using non-simultaneous photometry.

\begin{table}[!b]
\caption{Trace on color magnitude diagram }
\centering
\renewcommand{\arraystretch}{1.2}
\label{tab:slope}
\begin{tabular}{lccccc}
\hline
\hline
  Mechanism & $\Delta B$ & $\Delta V$&$\Delta R$ & $\Delta I$ & $\Delta I$ / $ \Delta (B-I)$ \\
  \hline
  Spot   & 0.38 & 0.32 & 0.25 & 0.24 & 1.71 \\
  Accretion   & 0.20 & 0.07 & 0.04  & 0.02 & 0.11 \\
  Extinction dip $^{*}$   & 1.56 & 1.22 & - &0.60 & 0.63 \\
  Long term  & 1.80 & 1.50 & - &0.80 & 0.80 \\
  \hline
  \hline
\end{tabular}
\begin{flushleft}
$*$ The extinction dip represents the extinction event centered at Day 440.
\end{flushleft}
\end{table}

\section{Discussion}

Photometric dips, accretion bursts, and a 7.03 d periodicity all shape the light curve of GI Tau during our monitoring over two years. The properties of the inner edge of circumstellar disk and the star-disk interactions can be determined from the morphology and color changes during the variation events.  The existence of  quasi-periodic extinctions in the first year and the non-detection during our second campaign, and the change in morphology and frequency of events within each campaign, indicate an evolution of the inner disk structure over at most a few orbital timescales.  In this section, we discuss the 2014 -- 2015 quasi-periodicity in terms of a warp model, the extinction curve, and the distribution of accretion rates.

\subsection{The slow warp model for the quasi-periodic dips of 2014 -- 2015}

\begin{figure}[!t]
\includegraphics[width=3.4in,angle=0]{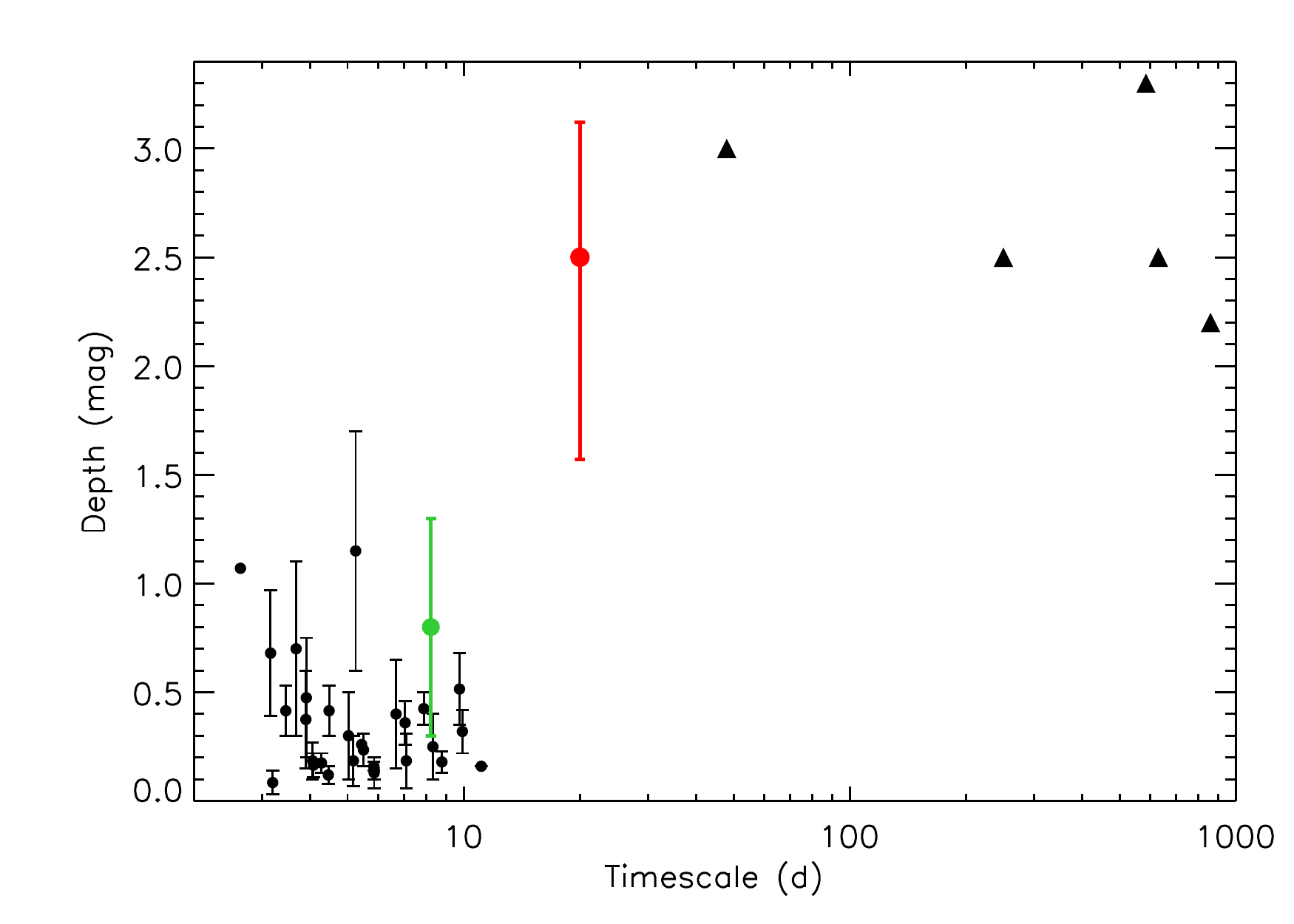}
\caption{The depth and timescale for extinction events of classical T Tauri stars, with the 2014--2015 quasi-periodicity and the months-long extinction from late 2015 shown in red.  Periodic or quasi-periodic targets from \citet{McGinnis2015, Stauffer2015, Ansdell2016} are shown as circles and cluster at periods consistent with stellar rotation and extinctions of 0.1--1 mag. Periodic variation of AA Tau is marked by green. Long-term extinction events of the faders KH 15D, RW Aur, V 409 Tau, and DM Ori from \citep{Kearns1998, Rodriguez2015,Rodriguez2016} are triangles and plotted with ``timescale'' indicating the duration of the event.  These extinction events are usually deeper, though this may be an observational bias.}
\label{fig:pvd} 
\end{figure}

Emission from young stars is periodically occulted by the inner edge of the circumstellar disk, when the disk is viewed close to edge-on. The presence of asymmetric disk warps or puffed-up inner rims will extinct the stellar brightness \citep[see e.g., the radiative transfer simulations of][]{Kesseli2016}.  Figure \ref{fig:pvd} presents the periods and amplitudes of extinction events seen on young stars.  For most dippers, these occultations are thought to occur once per stellar period, last $\sim 1$ day, and are caused by inner disk warps related to accretion funnel flows \citep[e.g.][]{Bouvier2007,Romanova2013}.  For faders, the occultations are prolonged and may last months or even years.  The GI Tau lightcurve exhibits some characteristics of both dippers and faders. 

In 2014 -- 2015 monitoring, the (quasi)-periodic dips of $1.5-2.5$ mag in $V$ occurred every $\sim 21$ days.  In contrast, all previous periodic dippers have periodicity on much shorter timescales that are consistent with the stellar rotation period \citep{Grankin2007, Bouvier2007, Alencar2010, McGinnis2015} and have depths of $A_V=0.1-1$ mag.  The deep obscuration depth of GI Tau in this campaign is comparable to UXors, which are usually early type PMSs undergoing variable extinctions with depths $A_V > 1$ mag \citep{Grinin1991, Grinin1994, Herbst1994, Natta1997, Dullemond2003}. However, no clear periodicity has been reported on UXors. 

The deep events of GI Tau recur near every $\sim 3$ stellar rotation periods and may be evidence of the slow warp in the MHD simulations of magnetospheric accretion by \citet{Romanova2013}.  In these simulations, two warps form in the circumstellar disk:  a thin warp located at the co-rotation radius ($R_{\rm cor}$) and a thick warp outside of the co-rotation radius.  Material can be trapped by the thick warp because of coupling between the stellar rotation and global oscillations in the disk.  The thick warp is expected to rotate several times more slowly than the star, since it is located at a larger radii in the disk, and also cause dips that are more optically-thick than thin warps at the inner disk edge.  The thick warp has a high scale height, so that it periodically intercepts our line-of-sight and causes extinction.  Although this slow warp was quasi-periodic over $\sim$60 days, the feature was short-lived:  it formed soon after our initial 20-night monitoring and had evolved  or dissipated by the next year.

The $\sim$80 day-long fade and return at the end of 2015 is much shorter than equivalent events on other stars, such as the years-long fading on AA Tau and V409 Tau \citep{Bouvier2013, Rodriguez2015}.  The obscuration source may be an azimuthally symmetric warp located close to the inner edge of the disk \citep[e.g.][]{Dullemond2003}, distant disk structures \citep[e.g.][]{Zhang2015} or a bridge between an outer and inner disk \citep{Loomis2017}.

As the obscuration source of the extinction dips is located not far from the inner edge of circumstellar disk or co-rotation and truncation radius, we calculate the co-rotation radius of GI Tau based on the stellar parameters, and spin period obtained from this work.
\begin{equation}
R{_{\rm cor}} = ({\rm G M_* P_*^2})^{1/3} (2\pi)^{-2/3} = 7.35 R_* = 0.06 \rm \, AU , 
\end{equation}
where $\rm M_* = 0.53 M_{\odot}$, $\rm R_* = 1.7 R_{\odot}$, and $\rm P_* = 7.03$ d. 

The morphology of the dips is related to the disk inclination, orientation of magnetic field dipole, and warp opacity. The short-durations of the dips detected on GI Tau indicates a moderate inclination viewing angle \citep{Bodman2017}. The shape of the dips depends on the ingress timescale, i.e., the timescale for the structure to move in front of the star.  The orbital velocity is calculated by the duration of the ingress time following the equation: 
\begin{equation}
V_{\rm orbit} \times \rm sin\, \theta = L / t_{ \rm ingress}
\end{equation}
\noindent where the definition of $L$ is half of the angular size of the warp \citep{Bouvier1999}, and the $t_{ \rm ingress}$ should around half of the total obscuration time. As shown in Figure \ref{fig:gi389}, the typical $t_{ \rm ingress}$ is 4 days while the occultation last for 8 days.  A disk warp located at $\sim 1.5$ R$_{\rm cor}$ has a local disk rotation velocity $v_{\rm rot} = $ 43.5 km/s. A Gaussian shape warp modeled by \citet{Romanova2013} with $v_{\rm warp} = 0.25 \, v_{\rm rot}$ should have a width L =  $6.9 \, R_*$ in horizontal size for an 8-day duration.

The maximum observed duration of the dips in 2014 - 2015 campaign is 5 days, or 25\% of the occultation period ($P \sim 20$ days).  If we assume the warp system is located at 1.2 to 1.5 co-rotation radius, as indicated by the \citet{Romanova2013} simulations, the angular width of the warp $L$ is as large as 2.35 $R_{cor}$ or $\sim$18.6 $R_*$. A hydrogen gas column density is derived by \citet{Bohlin1978}: $N_{\rm H}/E(B-V)$=$5.8\times 10^{21}$ cm$^{-2}$ mag$^{-1}$, assuming a $R_V = 3.85$ extinction (See \S 4.2).    We also assume an ISM gas-to-dust ratio as $100:1$, although this may not be valid for inner disks.
The gas mass within the warp is then roughly estimated by:
 \begin{equation}
 M_{\rm{warp,gas}}=1.5 \times 10^{21}\times A_V \times m_{\rm H} \times S_{\rm{warp}},
 \end{equation}
 \noindent where $m_{\rm H}$ is the atomic mass of hydrogen and $S_{\rm{warp}}$ represents the cross-section area of warp. We infer from the lightcurve that the warps have a Gaussian shape with a central height $H = 2 \, R_*$. The estimated gas mass is $1.6 \times10^{20}$ g for warps with an average extinction of $A_{V} = 1$ mag. The short-duration extinction events in 2015 -- 2016 are less deep and would therefore either have less mass or a lower scale height.

\begin{figure}[!t]
\includegraphics[width=3.3in,angle=0]{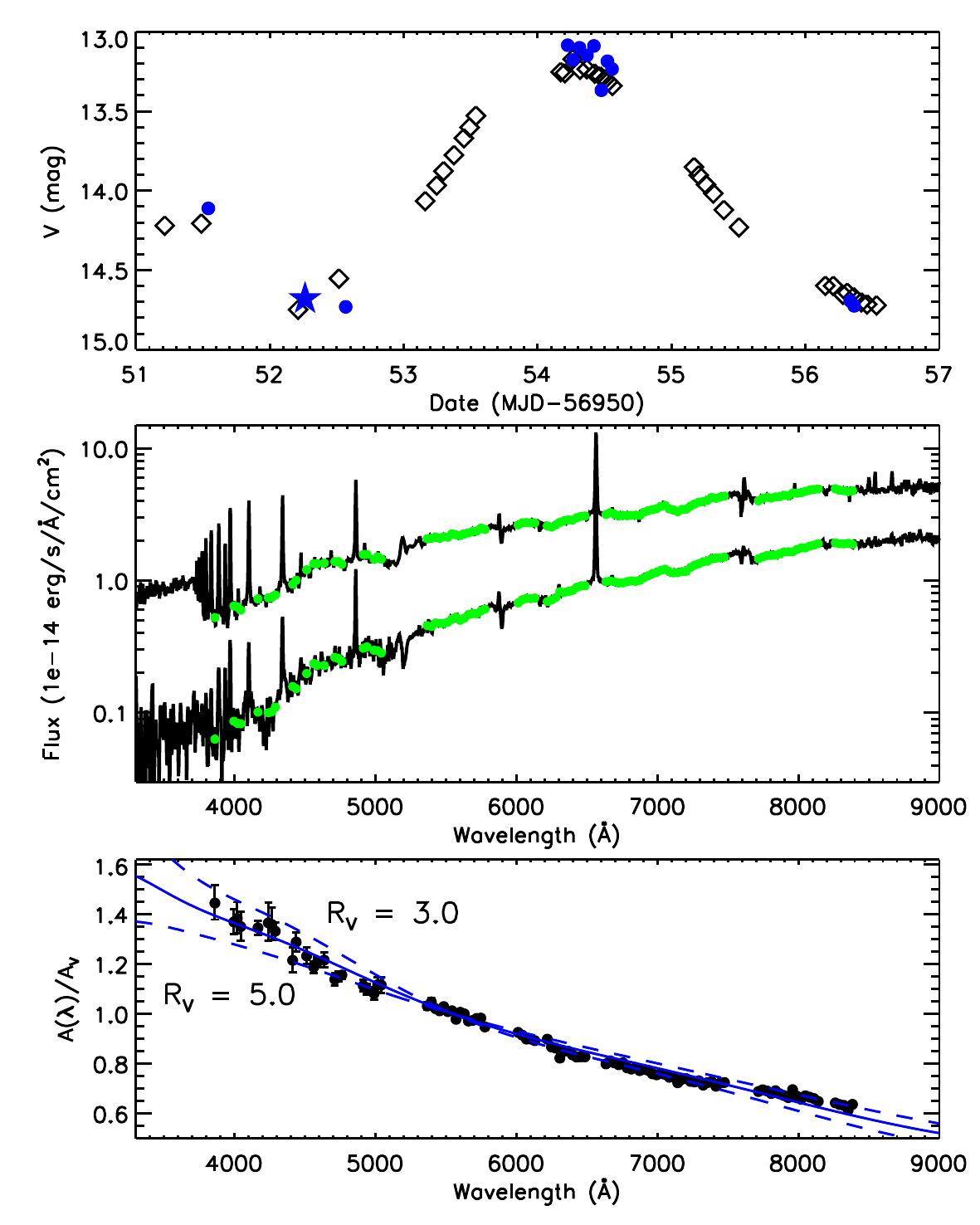}
\caption{{\it Top:} The Bessell $V$-band lightcurve of GI Tau during the SNIFS survey. The photometry by acquisition images are shown by black diamonds, while the blue dots and stars are synthetic photometry obtained from our flux-calibrated spectra.  {\it Middle:} Two SNIFS spectra of GI Tau, with one obtained during a bright epoch on Day 54 and one obtained during a faint epoch on day 52 (both marked as stars above).  The green dots mark the locations of the spectra used to measure the extinction law.  {\it Bottom:} The extinction law (flux ratio) of the spectra shown in the middle panel, normalized by $A(\lambda)$ at 5500 \AA. The blue lines show reddening curves of \citet{Cardelli1989} for $R_V = 3.85$ (solid) and 3.0 and 5.0 (dotted).}
\label{fig:rv} 
\end{figure}

\subsection{The Extinction Curve of the Dips of GI Tau}

Extinction events in single-band photometry have degenerate explanations: the star may be entirely occulted by dust described by some column density and extinction law, or a fraction of the star may be entirely occulted by a large column of dust \citep[see discussion in, e.g.][]{Bodman2017}.  If the star is entirely occulted by dust, then the wavelength dependence of the extinction will lead to an estimate of grain growth, as long as reflected light is not significant.  If only a fraction of the star is covered by opaque dust, then the star will get fainter but the color will not change.

Figure~\ref{fig:rv} shows flux-calibrated spectra of GI Tau obtained at minimum brightness during an extinction event and maximum brightness obtained at the end of that event\footnotetext{The near-simultaneous photometry is 0.1 mag fainter than the synthetic photometry obtained from our flux-calibrated spectra.  However, the $\Delta V$ measured from the synthetic spectra are consistent with the $\Delta V$ from photometry}.  The ratio of the two spectra demonstrate that GI Tau is much redder during occultation than out of occultation.  The TiO band ratios and Balmer Jumps are similar, indicating that the changes are caused by extinction rather than any change in spot coverage or accretion.  The redder spectrum in this epoch is consistent with our other spectra obtained during the same run, the few spectra analyzed by \citet{Herczeg2014}, and our photometric results.

The flux ratio between 4000--8500 \AA\ is fit with an extinction curve from \citep{Cardelli1989}, with free parameters $A_V$ and a total-to-selective extinction $R_V$ between 2.1 -- 5.8.  The best-fit $R_V=3.85\pm0.5$ indicates possible grain growth relative to the ISM. This fit is constrained primarily by flux at $<5000$ \AA.  The flux ratio \footnote{The flux ratio does not include any jump at 8200 \AA\ that could be caused by Paschen absorption in the gas in our line-of-sight.} of the spectrum deviates from the fit above 8000 \AA\ for all $R_V$.  This analysis ignores any contribution from dust scattering, which is likely important at bluer wavelengths \citep[see, e.g., analysis of AA Tau by][]{Schneider15aa}.  The $V$-band magnitude of the fainter spectrum is in the range where the ``blue turnaround" makes the spectrum appear bluer than one would expect from extinction alone. If considered, scattering would lead to a lower $R_V$ and may also explain the deviation at red wavelengths.  If some fraction of the star is covered by a much higher dust extinction, then $R_V$ would need to be much lower for the visible fraction of the star. 

Diffuse interstellar bands \citep[see review by][]{herbig95} are not detected in any spectrum, but would be expected to be strong if the dust composition were similar to the ISM \citep{friedman11}.  These bands are strong in lines-of-sight through molecular clouds \citep[e.g.][]{vos11}, and when seen in the spectra of some young stars \citep[e.g.][]{oudmaijer97,rodgers02} are likely caused by the interstellar medium rather than the disk.   Dust heating and processing within the disk of GI Tau must have destroyed the complex molecules that cause these bands.  This difference could provide a method to distinguish disk extinction from interstellar extinction.

The flux in the [\ion{O}{1}] 6300 \AA\ emission does not change between epochs, despite the change in extinction.  High-resolution spectra of GI Tau includes broad and narrow components \citep[e.g.][]{simon16}.  The bulk of this emission must originate above the star, where the outflow would not be occulted by a inner disk warp.

\begin{figure*}[!t]
\centering

\includegraphics[width=3.1in,angle=0]{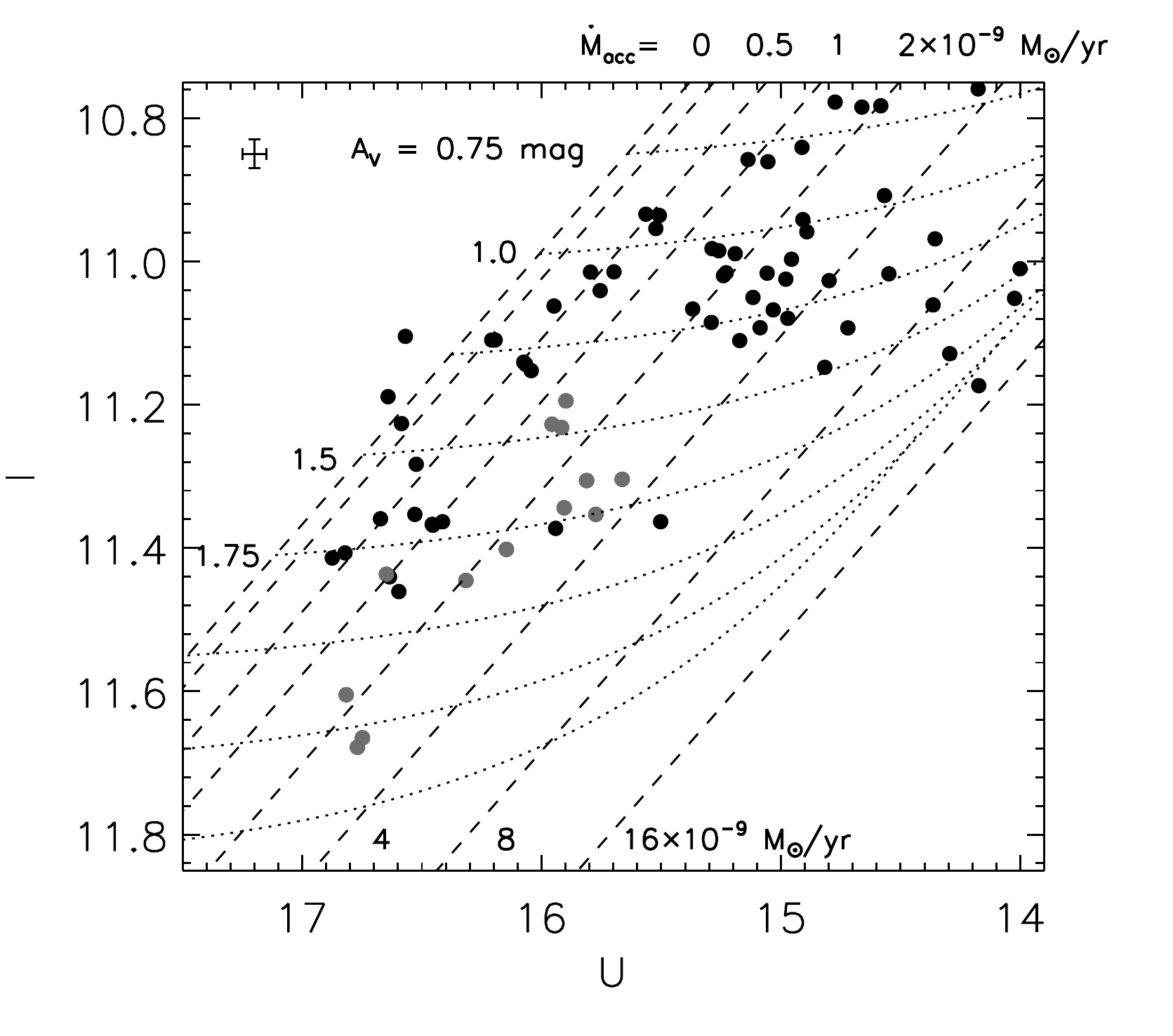}
\includegraphics[width=3.1in,angle=0]{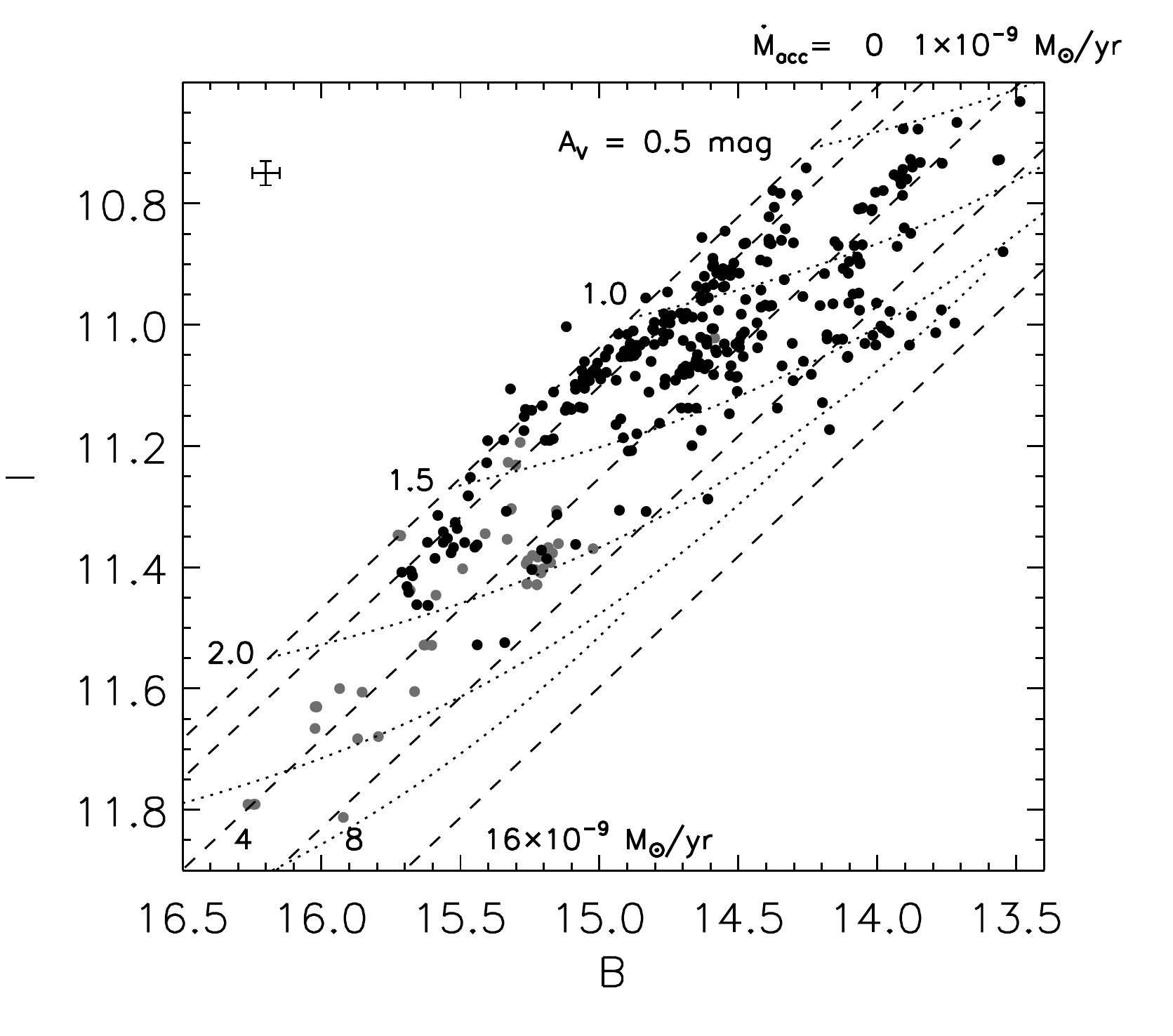}

\caption{The spot-corrected $U$ versus $I$ ({\it left}) $B$ and versus $I$ ({\it right})  during our 2015 -- 2016 monitoring of GI Tau.  The spots are removed as sinusoidal lightcurves with parameters in Table \ref{tab:sinfit}. The horizontal lines indicate accretion rates for the same extinction, while the diagonal lines indicate the extinction for the same accretion rate.  
This grid is calculated based on two assumptions: a) $I = U - 0.15$ as the accretion and b) extinction amplitudes in each band follow the $R_V = 3.85$ curve from \citet{Cardelli1989}. The estimated extinction ranges from $A_V = 0.5$ to $2.5$ mag assuming out of extinction brightness $I = 10.43$ mag \citep{Herczeg2014}.}
\label{fig:ui} 
\end{figure*}

The wavelength-dependent ratio of the two spectra is consistent with the other spectra obtained during the rise from day 52 -- 54.  The Balmer jump and therefore the accretion changes between days 54 -- 56, so the later spectra are not immediately useful for $R_V$ calculations.  On the other hand, when calculated from our photometry of extinction events (see Table \ref{tab:slope}), we obtained $R_V = A_V / (A_B - A_V) \sim 5$ for the long-term extinction (fader), and the dip on Day 440 (dipper) yields $R_V = 3.6$. The fits to the long-term fade may be less reliable because they include different points for each band and cover accretion bursts and spot rotation.

The $R_V$ measurement indicates a low opacity of the obscuration source, in contrast to previous interpretations that the periodic dips of AA Tau are optically-thick \citep{Bouvier2003}.  Any optically-thin dust in the accretion flow or at the inner disk edge should be quickly destroyed by strong stellar irradiation.  In MHD simulations, the accretion stream drags dust grains from the optically-thick disk \citep{Romanova2003}, which may replenish the dust in our line-of-sight.  However, the occultation timescales of the dips (e.g. 5 days) are relatively long compared with the crossing-timescale of an inner disk warp at the co-rotation radius.  Alternative explanations that the dust is located in disk winds at larger radii , rather than the disk itself, could explain the long survival time of the dust \citep{Bans2012,Petrov2015,Petrov2017}.

\begin{figure}
\includegraphics[height=2.9in,angle=0]{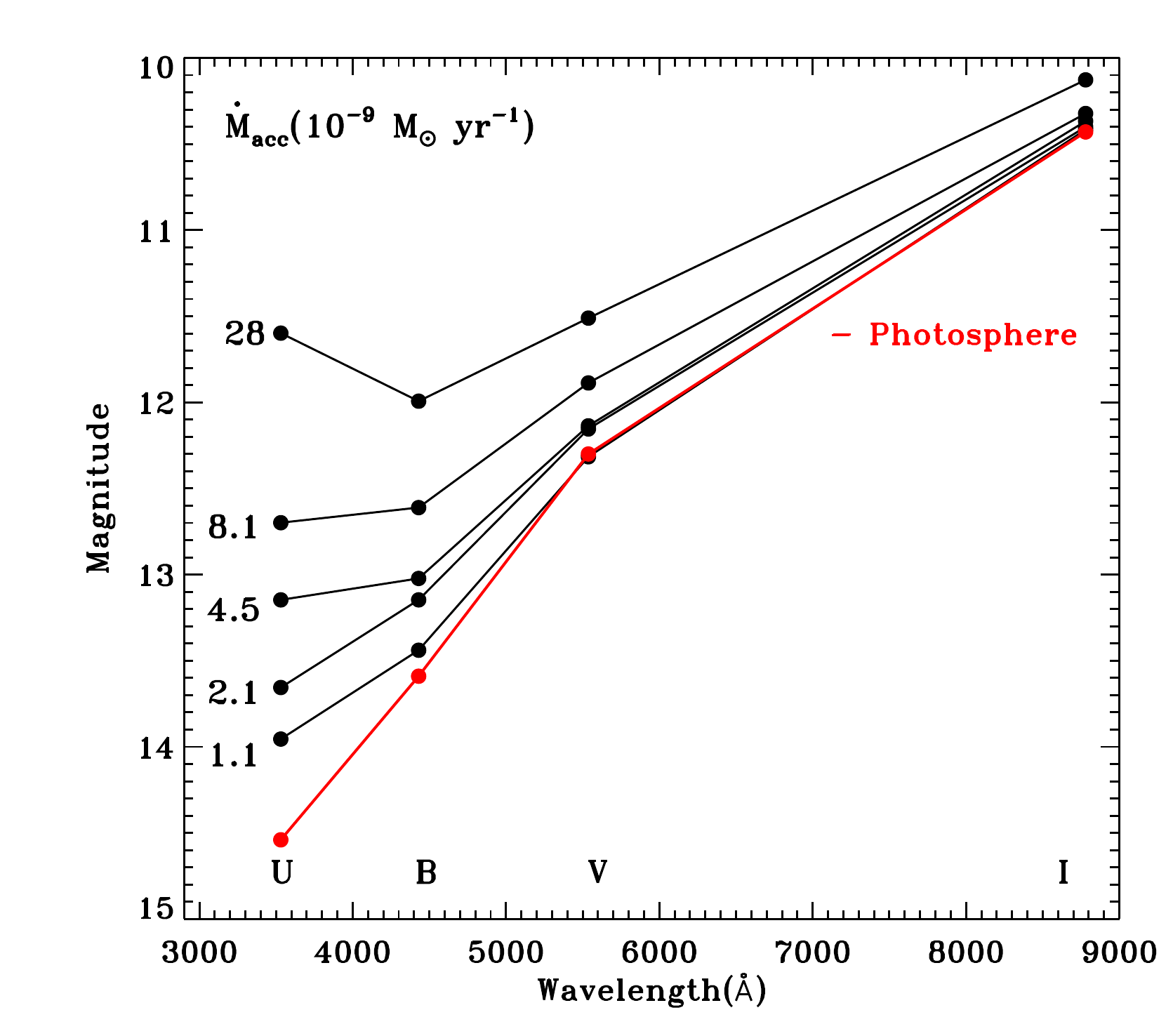}
\centering
\caption{The optical spectral energy distributions of GI Tau obtained at five different accretion rates, alongside a photospheric template (red). The photometry has been corrected for extinction.  The photospheric template is $U_{\rm photosphere} = 14.54$ mag, $B_{\rm photosphere} = 13.59$ mag, $V_{\rm photosphere} = 12.29$ mag and $I_{\rm photosphere} = 10.43$ mag. }
\label{fig:sed} 
\end{figure}

\subsection{Accretion on different timescales}

Mass accretion rates ($\dot{M}_{\rm{acc}}$) are measured here by calculating the excess continuum and line emission produced by the accretion flow.  Our $B$-band and limited $U$-band monitoring of 2015 -- 2016 are shown in Figure \ref{fig:gitau2015}, with variations caused by changes in accretion, extinction, and spot coverage.  Because scattered light during deep extinction events strongly affects the colors (the `blue turnaround'), accretion rates are calculated only for epochs when $V < 14.0$ mag.

To measure the excess $U$-band luminosity, we first remove the spot modulation effects by a 7.03-day sinusoidal lightcurve.  We then extract the extinction-corrected photospheric emission from the flux-calibrated optical spectra of \citet{Herczeg2014}.  The combined fit of a photospheric template and accretion continuum to the spectrum yields photospheric luminosities of $U_{\rm photosphere} = 14.54 \pm 0.1$ mag,  $B_{\rm photosphere} = 13.44\pm0.05$ mag, and $I_{\rm photosphere} = 10.43 \pm 0.05$ mag, when corrected to $A_V=0$ mag.   Any extinction-corrected $U$-band emission above this brightness is attributed to accretion. The color of accretion is calculated as $U - I \sim 0.15$ mag, using assumptions for the accretion continuum from \citet{Herczeg2014}, as estimated from veiling measurements of \citet{fischer11}. The color variations are then calculated for a variable extinction, following the $R_{V} = 3.85$ curve from \citet{Cardelli1989} with $A_U = 1.47 \, A_V$, $A_B = 1.25 \, A_V$, and $A_I = 0.56 \, A_V$. 
Figure~\ref{fig:ui} shows how extinction and accretion affect the $U-B$- and $I$-band magnitude of GI Tau.

The optical spectral energy distributions of spot and extinction removed examples are presented in Figure \ref{fig:sed}.   The accretion excess usually contributes $\sim 60$\% of the emission in the $U$-band filter but only $\sim 15$\% of the emission in the $B$-band filter on median mass accretion rate $\dot{M}_{\rm acc} = 1 \sim 4 \times 10^{-9} $ M$_{\odot} \rm yr^{-1}$, consistent with expectations from accretion models \citep[e.g.][]{Calvet1998}.  A similar relationship is seen by comparing the left and right panels of Figure~\ref{fig:ui} where the data points are more scattered in $U$.

\begin{figure}[!t]
\includegraphics[height=2.9in,angle=0]{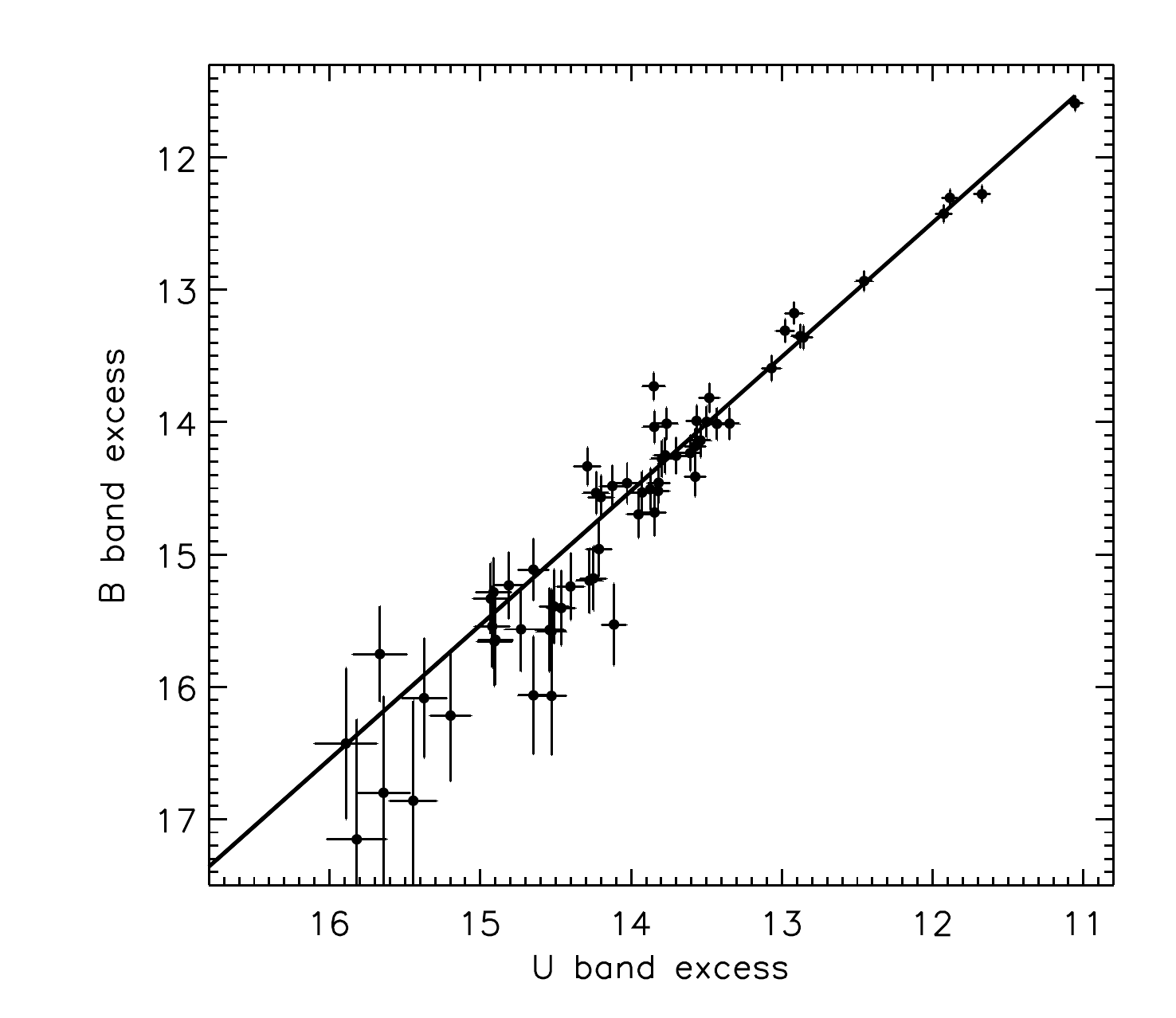}
\centering
\caption{The correlation of $U$- and $B$-band excess of GI Tau, both generated by accretion. The photometry has been corrected for spots, de-reddened, with an excess then measured against an estimated photospheric magnitude of $U_{\rm photosphere}=14.54$ mag; $B_{\rm photosphere}=13.44$ mag. The best linear fitting result is: $U_{\rm ex} = 0.93 B_{\rm ex} + 0.52$.}
\label{fig:ub} 
\end{figure}

\begin{figure*}
\centering
\includegraphics[width=2.8in,angle=0]{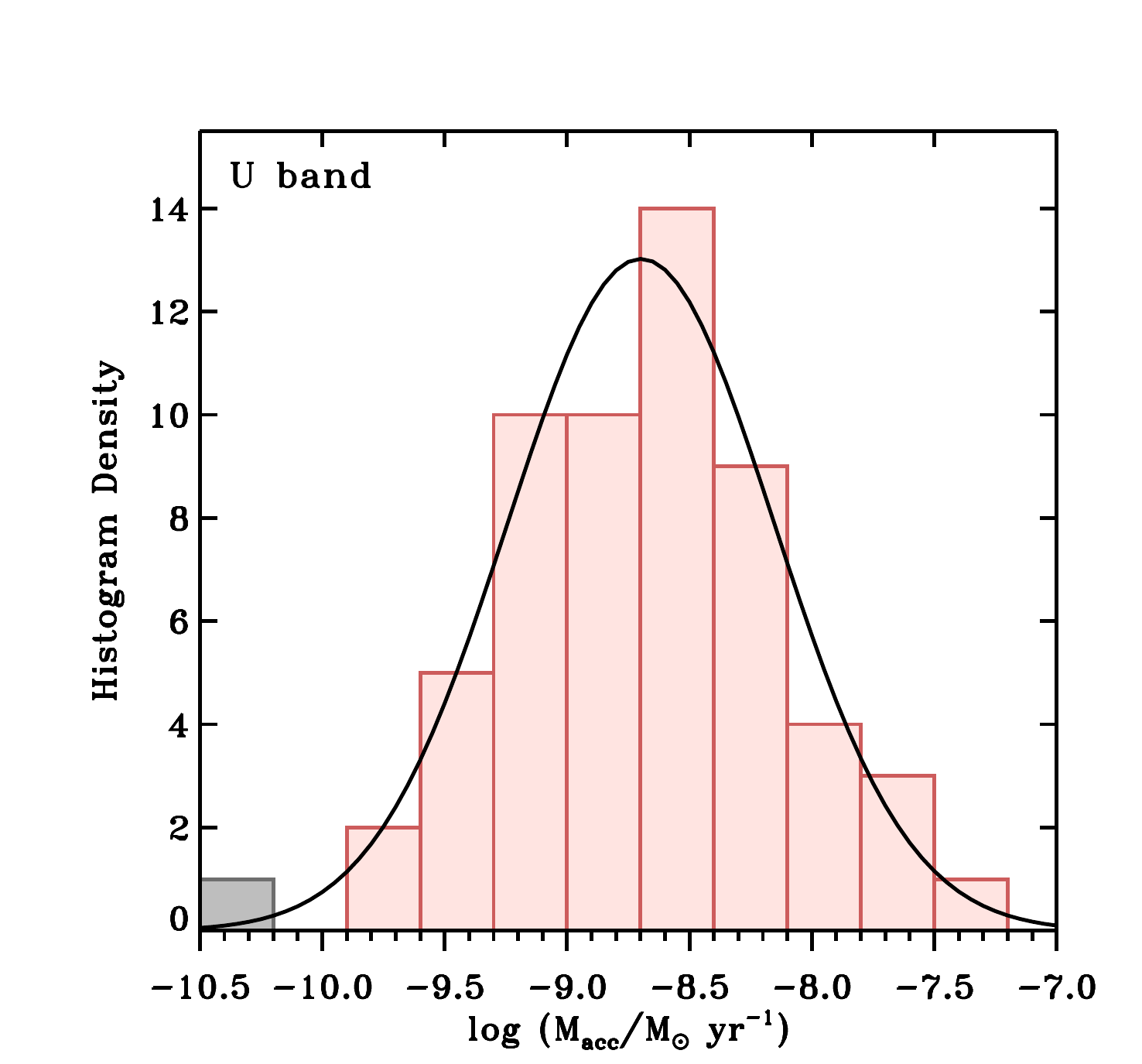}
\includegraphics[width=2.8in,angle=0]{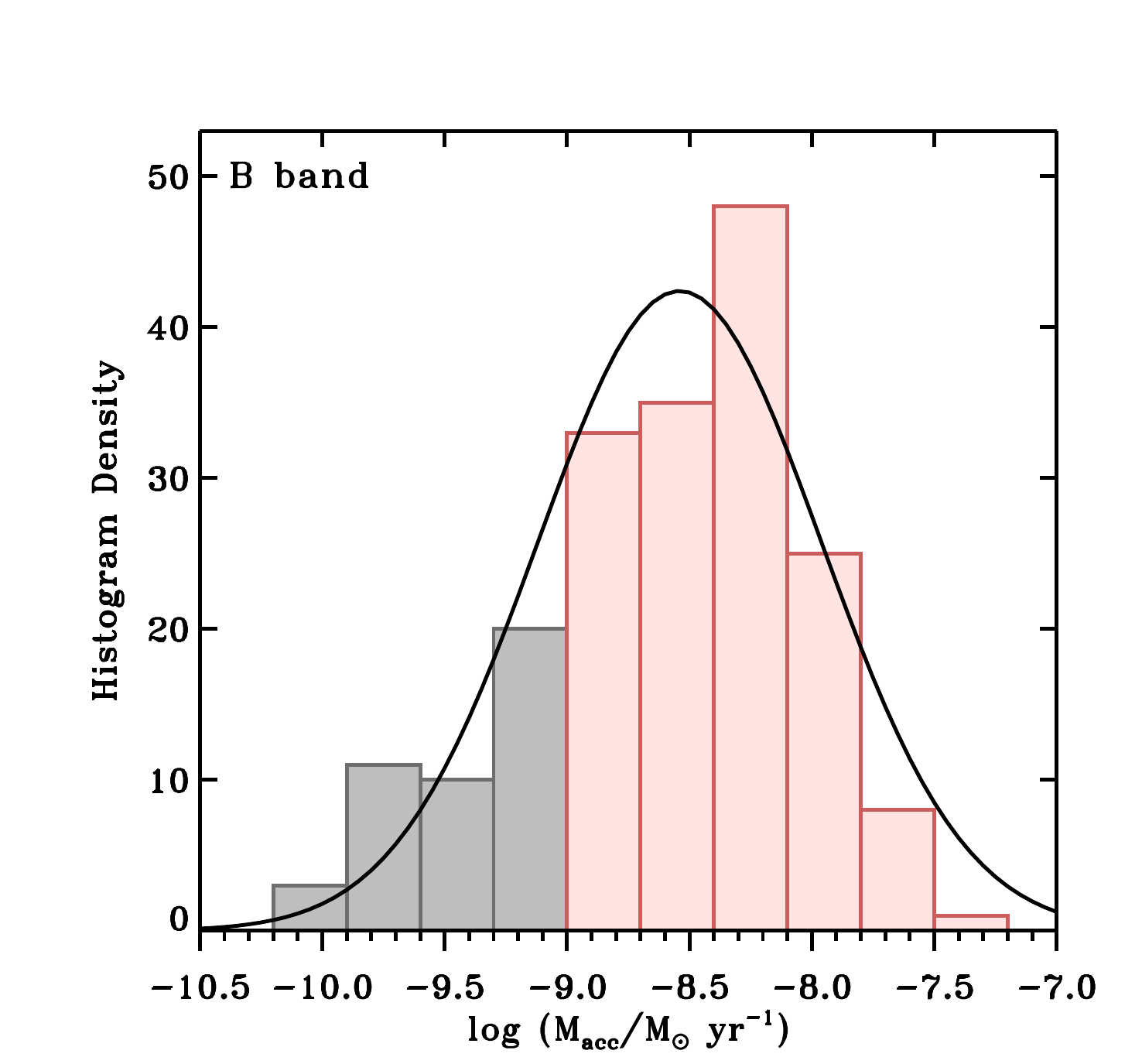}
\caption{ Histograms of accretion rates calculated by $U$ ({\it left}) and $B$-band ({\it right}) excess through the entire 2015 -- 2016 campaign. The data points taken within 2 hours are binned as one. The mass accretion rate higher and lower than the detection limit are shown by pink and grey, respectively. Gaussian fits of the histograms are shown by thick lines.}
\label{fig:acchis} 
\end{figure*}

Following the empirical relationship from \citet{Gullbring1998}, ${\rm log}(L_{\rm acc}/L_{\odot}) = \,1.09^{+ 0.04}_{- 0.18}\, {\rm log}({L^{\rm ex}_U/L_{\odot}})+\,0.98^{+0.02}_{-0.07}$, the accretion luminosity of GI Tau is calculated by the $U$-band accretion luminosity, $L_{\rm{acc}}$ by
\begin{equation}
L^{\rm acc}_U = 4 \pi d^2 F_{\rm zeropoint} \times (10^{-0.4U_{\rm unred}} - 10^{-0.4U_{\rm photosphere}} ),
\end{equation}
where $F_{\rm zeropoint}$ is the zero point of generic $U$-band, distance $d = 140$ pc, and $U_{\rm unred}$ is spot modulation and extinction reddening removed $U$ magnitude.  The accretion luminosity ranges from $\sim 0$ to $41 \times 10^{-2}$ L$_{\odot}$. 
 The accretion rate $\dot{M}_{\rm{acc}}$is then derived from the accretion luminosity, 
\begin{equation}
 \dot{M}_{\rm{acc}}\sim 1.25 L_{\rm{acc}}R_*/GM_*,
\end{equation}
\noindent where $R_{*}$ and $M_*$ are radius and mass of GI Tau. The calculated mass accretion rate of GI Tau ranges from $\sim0-52 \times 10^{-9}$ M$_{\odot}$yr$^{-1}$, for stellar parameters $R_* = 1.7$ R$_{\odot}$ and $M_* = 0.53$ M$_{\odot}$. 

We also develop a method to estimate accretion rate from $B$-band photometry, because our time coverage in $B$ is more extensive than in $U$.   After removing the sinusoidal spot modulation, the extinction and accretion for each $B$ and $I$ data point is estimated from the grid shown in Figure~\ref{fig:ui}.  The excess $B$-band emission produced by accretion is calculated by
\begin{equation}
B_{\rm ex} = -2.5 {\rm log}(10^{-0.4B_{\rm unred}}-10^{-0.4B_{\rm photosphere}}),
\end{equation}
where $B_{\rm unred}$ is the de-reddened magnitude in $B$-band using extinction curve of $R_V = 3.85$.  Figure~\ref{fig:ub} shows a linear relationship between nearly-simultaneous $U_{\rm ex}$ and $B_{\rm ex}$, with a best-fit
\begin{equation}
U_{\rm ex} = 0.93 B_{\rm ex} +0.52.
\label{eq:ub}
\end{equation}
The bolometric correction of $B$-band excess is then combined with Equation \ref{eq:ub} and the empirical relationship given by \citet{Gullbring1998}, as
\begin{equation}
{\rm log}(L_{\rm acc}/L_{\odot}) = \,1.22^{+ 0.05}_{- 0.19}\, {\rm log}({L^{\rm ex}_B/L_{\odot}})+\,1.46^{+0.06}_{-0.10}.
\end{equation}
Based on the accuracy of our photometry and the correction for spots, estimated as $\sim 0.1$ mag in both $B$ and $U$-band, our detection limits of accretion rate measurement are set as $\log \rm (M_{\rm acc}/M_\odot yr^{-1}) > - 9.0$ for $B$-band and $> - 10.0$ for $U$-band. The correlation between near-simultaneous $B$-band and $U$-band accretion rates is tight at rates higher than $\log \rm (M_{\rm acc}/M_\odot yr^{-1}) > - 8.2$ but unreliable at lower accretion rates.

The mass accretion rates of GI Tau calculated from $U$ and $B$-band excesses are summarized in Figure~\ref{fig:acchis}.  
As measured from the U-band excess, the 5th to 95th percentile range of $\log \rm (M_{\rm acc}/M_\odot  yr^{-1})$ is $-7.89$ to $-9.77$, with a center of $-8.70$ and sigma as 0.53 dex in the Gaussian fit.  These results are consistent with results from the more-extensive $B$-band photometry, which yielded an average $\log$ M$_{\rm acc}$/M$_\odot$ yr$^{-1}=-8.55$ with 0.6 dex scatter.  These estimates are obtained by creating mock sets of accretion rates over a range of values for the average and standard deviation and assuming a Gaussian distribution and upper limits.  The adopted values are then obtained from maximizing the probability from a Kolmogorov-Smirnov test between the observed distribution and each mock data set.  The distribution of $B$-band accretion rates includes the NOWT data sampled at a time-resolution of one hour.
The best-fit $B$-band data over predicts the number of data points at high accretion rates, as seen in Figure~\ref{fig:acchis}.  Differences in results between $B$-band and $U$-band accretion rates are likely attributed to the large scatter in $B$-band at average and weaker accretion rates.



\begin{figure}
\centering
\includegraphics[width=3.in,angle=0]{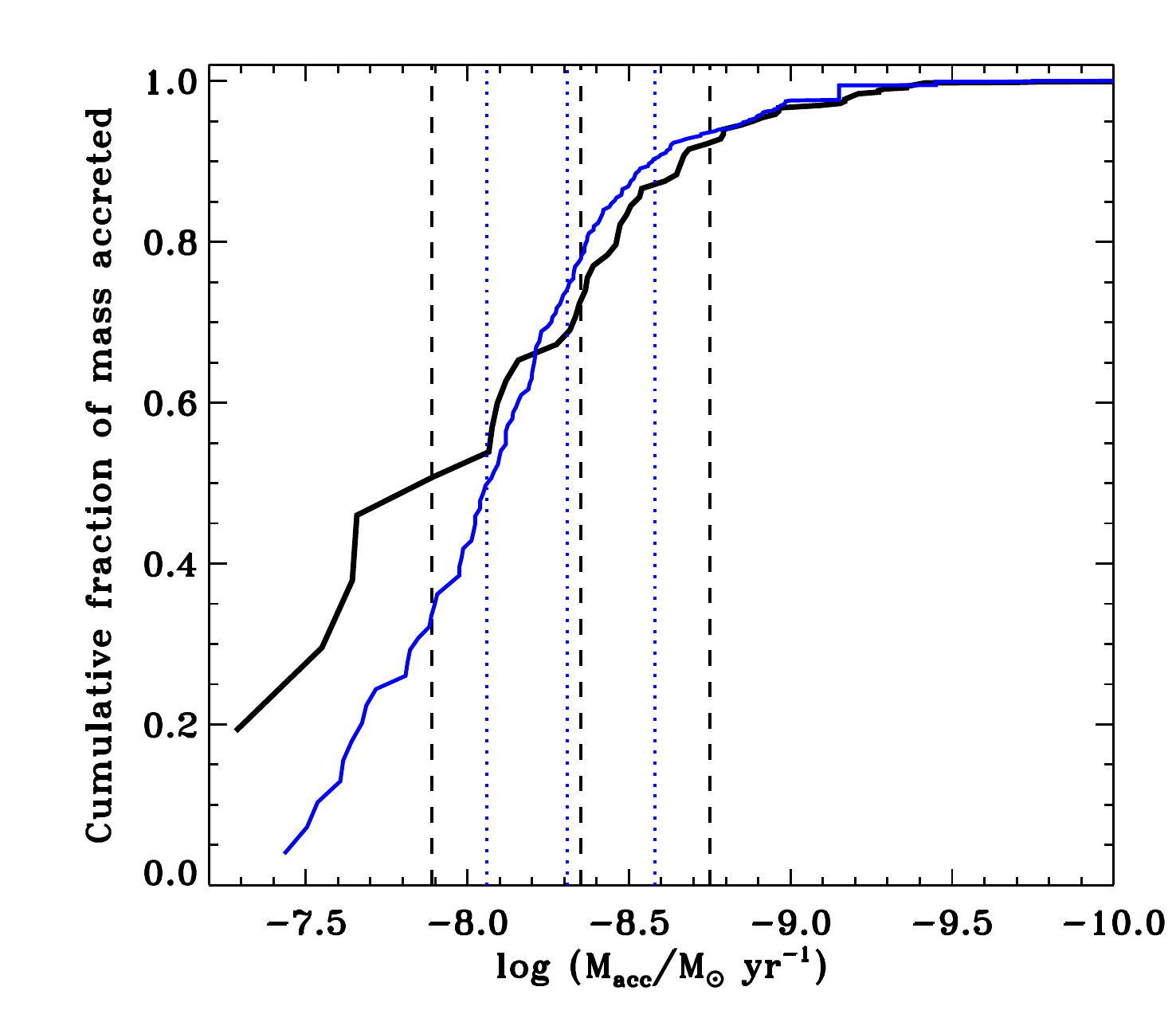}
\caption{The distribution of mass accretion rate measured by $U$ ({\it black}) and $B$ ({\it blue}) band photometry. Vertical dash/dot lines from left to right indicate the accretion rate above which half the mass is accreted, the average accretion rate, and the average mass accretion rate in log space.}
\label{fig:mdot} 
\end{figure}

This distribution of accretion rates is consistent with the distribution of accretion rates measured for stars of similar mass \citep[e.g.][]{Fang2013,Venuti2014,Manara2017}. However, the distribution demonstrates the importance of accretion bursts in models of disk evolution.  The average mass accretion rate of GI Tau is $4.7\times10^{-9}$ M$_\odot$yr$^{-1}$, two times faster than the average inferred from the $\log \rm (M_{\rm acc}/M_\odot  yr^{-1}) $.  Moreover, a total of 50\% of the mass is accreted when the accretion rate is higher than $12.8\times10^{-9} \,\rm M_\odot \rm yr^{-1}$, during accretion bursts (Figure \ref{fig:mdot}).  Such bursts are seen in our high-cadence NOWT monitoring, where for example  the accretion rate increased from $\sim 2.3\times 10^{-9}$ M$_\odot$yr$^{-1}$ to $7.3\times 10^{-9}$ M$_\odot$ yr$^{-1}$ in several hours on Day 458.

The periods of high accretion deplete most of the disk; the periods of low accretion are irrelevant. However, models of disk evolution \citep[e.g.][]{Rosotti2016,Rafikov2017,Mulders2017,Lodato2017} assume that the accretion rates are static.  Although these distributions cannot be fully explained by variability \citep{Costigan2014, Venuti2015}, and surely include some stars that are strong accretors and others that are weak, bursts should be expected to play a significant role in the mass accretion.  The distribution of high accretion rates could also be in excess over a Gaussian distribution. Future analyses should incorporate time-averaged accretion rates \citep[e.g.][]{Venuti2015} over many epochs and perhaps even many years.

\section{Conclusions}

Our two-year multi-band photometric monitoring of the classical T Tauri star GI Tau revealed variability caused by extinction, accretion, and spots, each with unique signatures in color-magnitude diagrams.  The deep extinction events of $\Delta V = 2 - 3$ mag were seemingly stochastic in their timing and duration, with some occultations lasting 3--5 days and one 80 day long dimming.  
During 3 months in 2014 -- 2015, the short dips recurred with a quasi-period of $\sim$ 21 days, as might be expected from the sub-Keplerian slow warp seen  in the simulations of \citet{Romanova2013}.  The stellar rotation period of $7.03\pm 0.02$ days is recovered from the second half of the 2015 -- 2016 lightcurve but is not apparent in our earlier lightcurve, consistent with previous period estimates from some epochs and with an inability to recover that period in other epochs.  

A wavelength dependent extinction curve is fitted by spectral ratios, with best fit R$_V$ = 3.85 $\pm$ 0.5. Diffuse interstellar bands are not detected from the spectra. The average mass accretion rate of GI Tau of $\sim4.7 \times10^{-9}$ M$_\odot$ yr$^{-1}$ is calculated from excess $U$- and $B$- band light curves, after accounting for extinction and spots.  The distribution of accretion rates demonstrates that most accretion occurs during bursts, so the quiescent accretion rates may provide a misleading evaluation of accretion as a diagnostic of disk physics.

\begin{table}
\centering
\caption{ Photometric period of GI Tau}
\renewcommand{\arraystretch}{1.2}
\label{tab:period}
\begin{tabular}{lccccc}
\hline
\hline
  Year       &Period (day) & Amp. $V$ (mag) & Number of Obs. & Ref\\
  \hline
  1984    & 7.18$\pm$0.05 & 0.22 & 68 & a\\
  1987    &7.13$\pm$0.06 & 0.34 & 38 & b\\
  1988    &7.01$\pm$0.17 & 0.33 & 45 & b\\
  1989    &7.00$\pm$0.06 & 0.20 & 66 & b\\ 
  1990    &7.06$\pm$0.05 & 0.35 & 57 & b\\ 
  1991    &7.28$\pm$0.18 & 0.40 & 31 & b\\
  1992   &7.33$\pm$0.14 & 0.47 & 24 & b\\
  1993   & -- & 1.64 & 35 & b\\
  2003  & -- & 0.60 & 9 & c \\
  2014   &$(21)$ & 2.20 & 174 & d\\
  2015    & 7.03$\pm$0.02 & 0.26 & 324 & d\\
\hline
 \hline
\end{tabular}
\begin{flushleft}
{Note: The periods listed in this table are photometric periods of GI Tau. In this work, we claim that the $\sim 7$ days periods are close to the stellar spin and the 21-day is an obscuration period contributed by `slow warp' located outside the inner edge of circumstellar disk. The Amp. $V$ here is the amplitude of sinusoidal fit from Generalized Lomb-Scargle (GLS) periodogram, and is not represent the obscuration depth. In the year 1993 and 2003, there is no period detected from the periodicity analysis. 
References: a) \citet{Vrba1986}, b) \citet{Herbst1994}, c) \citet{Grankin2007}, d) this work.}
\end{flushleft}
\end{table}

\section*{Acknowledgement}

We thank the anonymous referee for helpful comments and suggestions that improved the clarity and robustness of the results.  ZG thanks Prof. Douglas N. C. Lin for helpful discussions. We also thank Hiro Takami, Stefano Facchini, and Carlo Manara for discussions on RW Aur and Petr Petrov for discussions on extinction in disk winds.   We also thank all the observers and staff who contributed to this project, including those at HCT  (operated by Indian Institute of Astrophysics), YNAO, VBO, TNO, HCT, and Lulin observatories.  

We thank Guojie Feng, Chunhai Bai, Shuguo Ma, Guangxin Pu, Abudusaimaitijiang Yisikandeer and Xuan Zhang from Xingjiang Astronomical Observatory for organizing and running the NOWT observations that is partially supported by the CAS "Light of West China" program (2015-XBQN-A-02). 

ZG, GJH, and JJ are supported by general grant 11473005 awarded by the National Science Foundation of China.  JNF acknowledges the support from the National Natural Science Foundation of China (NSFC) through the grant 11673003 and the National Basic Research Program of China (973 Program 2014CB845700 and 2013CB834900).  

\bibliographystyle{apj}
\bibliography{GIpaper}

\end{document}